\documentclass[11pt,oneside,reqno]{amsart}

\usepackage{graphicx}
\usepackage{pict2e}
\usepackage{amssymb}
\usepackage{amsthm}                
\usepackage[margin=1in]{geometry}  
\usepackage{graphics}
\usepackage{epstopdf}
\usepackage{amsmath}
\usepackage{graphicx}
\usepackage{subfigure}
\usepackage{latexsym}
\usepackage{amsmath}
\usepackage{amsfonts}
\usepackage{amssymb}
\usepackage{tikz}
\usepackage{mathdots}
\usepackage{comment}
\usepackage{float}
\usepackage[mathscr]{euscript}
\usepackage{enumerate}
\usepackage{epsfig}
\usepackage { graphicx }
\usepackage { hyperref }
\usepackage { graphics }
\usepackage { graphicx }
\usepackage{booktabs}
\usepackage{wrapfig,lipsum,booktabs}

\usepackage{caption}

\usepackage{algorithm}

\usepackage{etoolbox}

\usepackage {eufrak}
\usepackage[utf8]{inputenc}
\usepackage{longtable}
\usepackage[lite]{amsrefs}

\renewcommand{\PrintDOI}[1]{\href{http://dx.doi.org/\detokenize{#1}}{doi: \detokenize{#1}}%
	\IfEmptyBibField{pages}{, (to appear in print)}{}}

\theoremstyle{definition}
\newtheorem{theorem}{Theorem}[section]
\newtheorem{lemma}[theorem]{Lemma}

\theoremstyle{definition}

\theoremstyle{remark}
\newtheorem{remark}[theorem]{Remark}
\numberwithin{equation}{section}

\numberwithin{equation}{section}

\newcommand{\para}[1]{\vspace{10pt}\noindent\textbf{#1.}}
\title{An Efficient Data Retrieval Parallel Reeb Graph Algorithm}
\author{Mustafa Hajij}
\address{Santa Clara University}
\email{mhajij@scu.edu}
\author{Paul Rosen}
\address{University of South Florida}
\email{prosen@usf.edu}

\date{}
\dedicatory{}

\begin{document}
	
\maketitle 
	
\begin{abstract}
The Reeb graph of a scalar function defined on a domain gives a topologically meaningful summary of that domain.  Reeb graphs have been shown in the past decade to be of great importance in geometric processing, image processing, computer graphics, and computational topology. The demand for analyzing large data sets has increased in the last decade. Hence the parallelization of topological computations needs to be more fully considered. We propose a parallel augmented Reeb graph algorithm on triangulated meshes with and without a boundary. That is, in addition to our parallel algorithm for computing a Reeb graph, we describe a method for extracting the original manifold data from the Reeb graph structure. We demonstrate the running time of our algorithm on standard datasets. As an application, we show how our algorithm can be utilized in mesh segmentation algorithms.    
\end{abstract}

\tableofcontents

\section{Introduction}

Recent years have witnessed extensive research in topology-based methods to analyze and study data~\cites{carlsson2009topology, carlsson2008persistent}. The popularity of topology-based techniques comes from the generality and the robustness of the techniques and their applicability to a wide range of areas. The \textit{Reeb graph}~\cite{Reeb1946} has been one of the most successful topological tools in data analysis and data understanding. The Reeb graph is a data structure associated with a scalar function defined on a manifold. It gives an efficient topological summary for the manifold by encoding the evolution of the connectivity of its level sets. Reeb graphs, and their loop-less version, contour trees~\cite{boyell1963hybrid}, are of fundamental importance in computational topology, geometric processing, image processing, computer graphics, and more recently, data analysis and visualization. Examples of Reeb graph applications include quadrangulation~\cite{hetroy2003topological}, shape understanding~\cite{attene2003shape}, surface understanding and data simplification~\cite{biasotti2000extended}, parametrization~\cites{patane2004graph, zhang2005feature}, segmentation~\cite{werghi2006functional}, animation~\cite{kanongchaiyos2000articulated}, feature detection~\cite{takahashi2004topological}, data reduction and simplification~\cites{carr2004simplifying, rosen2017using}, image processing~\cite{kweon1994extracting}, visualization of isosurfaces~\cite{bajaj1997contour} and many others.

The past decade has witnessed an increase of large geometric data on which a scalar field is defined. This has yielded several challenges for the time efficiency of computing topological structures on such data.   The parallelization of the utilized algorithms is a natural direction one should take in order to improve the computational-time efficiency. In this article, we introduce an efficient shared memory parallel algorithm to compute the Reeb graph from a scalar function defined on a triangulated surface with or without a boundary. In addition, our algorithm provides a fast method for retrieving the surface data from the constructed Reeb graph. The data consists of the Reeb graph as well as the map that goes from the Reeb graph back to the manifold, called the \textit{augmented Reeb graph}~\cite{harvey2010randomized}.  For this purpose, we define an explicit map that associates the Reeb graph data to its corresponding data on the manifold.  As an application, we show how the Reeb graph can be used to identify and calculate curves with certain homological properties on a surface. Finally, we show how the data retrieval aspect of our algorithm, along with the curves that are extracted from the Reeb graph structure, can be used for mesh segmentation and mesh parameterization.

\subsection{Prior Work and Contribution}
Reeb graph literature is vast and ranges from the computational accuracy of the graph to its applications in data analysis and visualization. We provide an overview here.

\para{Reeb Graph Algorithms} The first provably correct algorithm to compute a Reeb graph on a triangulated surface was presented by Shinagawa and Kunii in~\cite{shinagawa1991constructing}.  They computed the Reeb graph in $O(n^2)$ time, where $n$ in the number of triangles in the mesh. This time was later improved to $O(n \log(n))$ by Cole-McLaughlin et al.~\cite{cole2003loops}. 

Reeb graphs have also been studied for higher-dimensional manifolds and simplicial complexes. An algorithm for computing Reeb graph for a $3$-manifold embedded in $\mathbb{R}^3$ is proposed in~\cite{tierny2009loop}. The first Reeb graph algorithm on an arbitrary simplicial complex is given in~\cite{chiang2005simple}.  This algorithm can handle a non-manifold input, but its worse case time complexity is quadratic. Reeb graph for a varying scalar function is studied in~\cite{edelsbrunner2004time}. Other Reeb graphs algorithms can be found in~\cites{pascucci2007robust,doraiswamy2009efficient,Salman12,HWW10,doraiswamy2008efficient}. Approximate Reeb graphs algorithms can be found in~\cites{biasotti2000extended,hilaga2001topology}. However, such algorithms may lead to inaccurate results. Data retrieval from the Reeb graphs, also referred to as \textit{augmented Reeb graphs}, has also been studied, and some algorithms have been presented, for instance~\cites{ge2011data,tung2004augmented,biasotti2008reeb}. 

A loop-free Reeb graph, also called a contour tree, has been used extensively in data analysis and data visualization. Algorithms for computing such graphs can be found in~\cites{carr2003computing,raichel2014avoiding,chiang2005simple,van1997contour,tarasov1998construction}. Contour tree have been used for scientific visualization~\cite{pascucci2004multi}, volume rendering~\cite{weber2002topology}, terrain applications~\cites{besl1992method,gupta1995manufacturing}. Contour tree data retrieval is studied in~\cite{gueunet2016contour}. For a thorough introduction to the contour tree and its applications, the reader is referred to~\cites{carr2004topological,raichel2014avoiding} and the references within.

\para{Reeb Graph Generalizations} Reeb graphs have also been used to study and analyze point cloud data. The applications are numerous, including data skeletonization~\cite{ge2011data}, retrieving topological information from point data such as homology group computation~\cites{dey2011,ch}, locus cut~\cite{dey2009cut}, data abstraction~\cite{natali2011graph}, and recovering structural information of a scalar function on a point data~\cite{chazal2009analysis}. In the context of point clouds, a relatively recent construction named \textit{Mapper}~\cite{singh2007topological}  has received a lot of attention, as it generalizes both the Reeb graph and contour tree. Mapper has found numerous applications~\cites{lum2013extracting,nicolau2011topology,robles2017shape,rosen2018using} and has been studied from multiple  perspectives~\cites{carriere2015structure,dey2017topological,munch2015convergence,hajij2018mog}.

\para{Applications of the Reeb Graph} 
There is a rich literature in computer graphics regarding the use of the Reeb graphs. Reeb graphs have been used in mesh segmentation~\cite{xiao2003discrete}, shape similarity~\cite{tung2005augmented}, shape matching~\cite{mohamed2012reeb}, feature-extraction~\cite{bajaj1997contour}, surface reconstruction~\cite{biasotti2000surface}, extracting tunnel and handle loops of a surface ~\cite{dey2008computing}, removing tiny handle in an isosurface~\cite{wood2004removing} and shape matching~\cite{hilaga2001topology}. See also~\cite{biasotti2008reeb} for further applications of Reeb graph in computer graphics.

\para{Parallelization of Topological Structures} 
The demand to compute large data sets has increased in the last decade and hence the consideration of topological computations parallelization. Multiple attempts have been made in this direction, including multicore homology computation~\cite{lewis2014multicore}, spectral sequence parallelization~\cite{lipsky2011spectral}, distributed contour tree~\cites{morozov2012distributed,gueunet2016contour}, distributed merge tree~\cite{morozov2013distributed}, alpha complexes~\cite{masood2020parallel}, and distributed Mapper~\cite{hajij2017distributed}.

\para{Contributions} 
In this paper, we give a parallel Reeb graph algorithm on arbitrary triangulated mesh with and without a boundary.  We prove the correctness of our method using fundamental theorems in Morse Theory. Moreover, we discuss the performance results that compare our approach to a reference sequential Reeb graph algorithm~\cite{doraiswamy2008efficient}. We then show how we can use the Reeb graph to retrieve certain curves on the manifold. Finally, we utilize the data retrieval aspect of our algorithm and give an application to surface segmentation. Specifically, this article has the following contributions:

\begin{enumerate}

    \item We give an efficient parallel algorithm that computes the Reeb graph of a piece-wise linear function defined on a triangulated $2$-manifold with and without a boundary.
    
    \item  Our method can be used to retrieve the manifold data from the Reeb graph. In other words, given a point in the Reeb graph, we give an efficient method to retrieve the manifold data that corresponds to that point. This feature, as well as feature (1), makes our algorithm an augmented Reeb graph algorithm. 

    \item We show how the homological properties of a Reeb graph can be used to extract certain curves on a surface, and we utilize our algorithms to give a mesh segmentation algorithm.

    \item The algorithms presented here are easy to implement and require minimal memory storage.

\end{enumerate}

\section{Morse Theory and Reeb Graphs} 
In this section, we review the basic background needed in this paper. We start by reviewing the basics of Morse theory and Reeb graphs on smooth manifolds. Then we discuss the corresponding piece-wise linear version. For more details on Morse theory, the reader is referred to~\cites{matsumoto2002introduction,banyaga2013lectures}.

\subsection{Morse Theory}

Morse Theory is a tool from differential topology that is concerned with the relations between the geometric and topological aspects of manifolds and the real-valued functions defined on them. One of the primary interests in this theory is the relationship between the topology of a smooth manifold $M$ and the critical points of a real-valued smooth function $f$ defined on $M$. Intuitively, Morse theory studies the topological changes of the level sets of a real-valued smooth function as the height of $f$ varies. Morse theory was first introduced by Morse~\cite{morse1934calculus} for infinite dimensional spaces. A comprehensive introduction to Morse theory on finite-dimensional manifolds is given in~\cite{milnor1963morse}. Also see~\cites{matsumoto2002introduction,banyaga2013lectures}. Morse theory has been proven to be a very useful tool in computer graphics and geometric data processing and understanding. The theory was extended to triangulated $2$-manifolds by~\cite{critical1967}. Recently, Morse theory has found applications in global surface parameterization~\cite{guo2006meshless}, finding a fundamental domain of a surface~\cite{fair2004}, surface quadrangulation~\cite{dong2006spectral}, topological matching~\cite{hilaga2001topology}, implicit surfaces~\cite{stander1997guaranteeing}, surface segmentation~\cite{yamazaki2006segmenting}, spline construction~\cite{wang2009geometry}, and many other applications. 

Let $M$ be a compact and smooth $n$-manifold and let $I=[a,b]\subseteq \mathbb{R} $, where $a<b$, be a closed interval. Let $f:M \longrightarrow I$ be a smooth function defined on $M$. A point $x \in M$ is called a \textit{critical point} of $f$ if the differential $df_x$ is zero. A value $c$ in $\mathbb{R}$ is called a \textit{critical value} of $f$ if $f^{-1}(c)$ contains a critical point of $f$. A point in $M$ is called a \textit{regular point} if it is not a critical point. Similarly, if a value $c \in \mathbb{R}$ is not a critical value, then we call it a regular value. The inverse function theorem implies that for every regular value $c$ in $I$, the level set $f^{-1}(c)$ is a disjoint union of $n-1$ manifolds. In particular, when $n=2$, then $f^{-1}(c)$ is a disjoint union of simple closed curves. A critical point is called \textit{non-degenerate} if the matrix of the second partial derivatives of $f$, called the \textit{Hessian matrix}, is non-singular. A differentiable function $f:M\longrightarrow I$ is called \textit{Morse} if all its critical points  are non-degenerate, and all critical values are distinct. If the manifold $M$ has a boundary, i.e., $\partial M \neq \emptyset $, then we will also require two other conditions: (1)~$f^{-1}(\partial I) = \partial M$ and (2)~there are no critical points on $\partial M$. In other words, the boundary points in the interval $I$, the values in $\partial I$, are regular values for the function $f$. The \textit{index} of a critical point $x$ of $f$, denoted by $index_f(x)$, is defined to be the number of negative eigenvalues of its Hessian matrix. For instance, the Hessian of a scalar function on a smooth surface is a $2\times 2$ symmetric matrix. Hence, the index of $f$ on a critical point takes the values $0$, $1$, or $2$. In this case, an index $0$, $1$, or $2$ of a critical point of a function $f$ is nothing more than a local minimum, a saddle, or a local maximum for $f$, respectively.

\begin{figure}[!ht]
    \centering
    \includegraphics[scale=0.29]{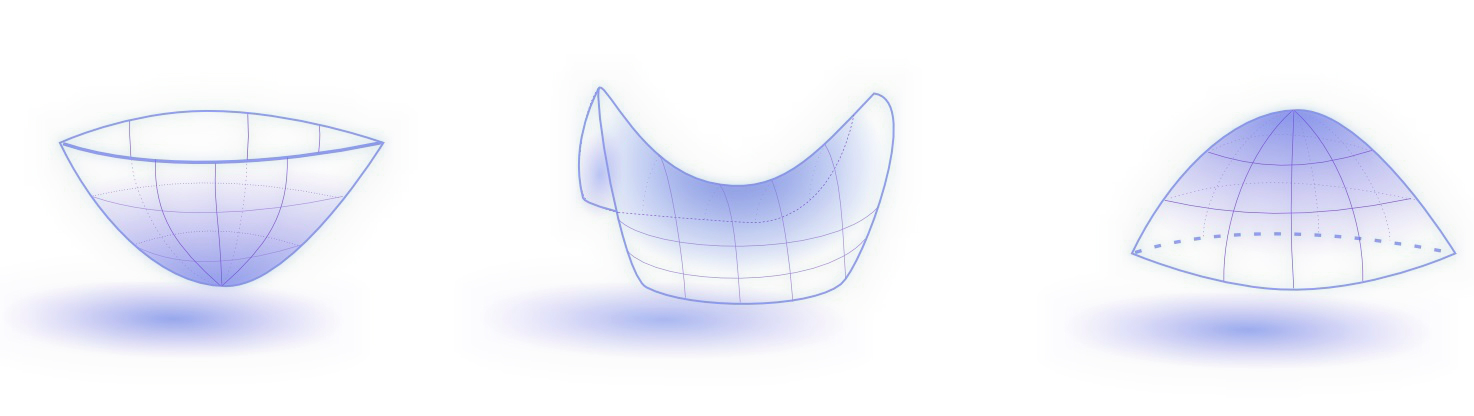}
    \caption{Minimum, Saddle, and Maximum, respectively.}
    \label{minmaxsaddle}
\end{figure}

If $f$ is a Morse function on a surface, then, up to a change of coordinates, the surface around a critical point $f$ has one of the simple forms as appear in Figure \ref{minmaxsaddle}. 
More formally, we have the following important Lemma:
\begin{lemma}
    (Morse Lemma) Let $M$ be a smooth surface, $f:M \longrightarrow \mathbb{R}$ be a smooth function, and $p$ be a non-degenerate critical point of $f$. We can choose a chart $(\phi,U)$ around $p$ such that $f \circ \phi^{-1}$ takes exactly one of the following three forms:
\begin{enumerate}

    \item Local minimum --- $f \circ \phi^{-1}(X,Y)= X^2+Y^2+c$
    \item Saddle --- $f \circ \phi^{-1}(X,Y)= X^2-Y^2+c$
    \item Local maximum --- $f \circ \phi^{-1}(X,Y)= -X^2-Y^2+c$
    \end{enumerate}
\end{lemma}

An analogous lemma holds for Morse functions on higher dimensional smooth manifolds. See~\cite{matsumoto2002introduction} for more details. Morse Lemma implies that a Morse function around a critical point looks simple, and it is exactly one of the forms given in the Lemma above, up to change of coordinates. Notice that the number of minus signs in the standard form of the function $f$ around $p$ is equal to the index of the critical point $p$.

\subsection{Reeb Graphs}
  
Let $M$ be a topological space and let  $f:M\longrightarrow \mathbb{R}$ be a scalar function defined on $M$. The Reeb graph of the pair $f$ and $M$ gives a summary of the topological information encoded in by tracking changes occurring to the connected components of the level sets of $f$. More precisely, the \textit{Reeb graph} of $M$ and $f$ is the quotient space $R(M,f)$ of $M$ defined as follows. We say the $x$ and $y$ are equivalent in $M$, and write $x \sim y$, if and only if they belong to the same connected component of $f^{-1}(r)$ for some $r\in \mathbb{R}$. The quotient space $M/\sim = R(M,f)$ with the quotient space topology induced by the quotient map $\pi: M\longrightarrow R(M,f)$ is called the Reeb graph of $M$ and $f$. Recall here that the map $\pi$ takes a point $x$ in $M$ to its equivalence class $[x]$ in $R(M,f)$. Given a point $p$ in $R(M,f)$, it is often important in practice to retrieve the set of points in $M$ that map to $p$ via $\pi$. In this article, we provide an efficient method to retrieve the data in $M$ associated with a point on a Reeb graph.

The map $\pi$ induces a continuous function $\bar{f}:R(M,f)\longrightarrow \mathbb{R}$, where $\bar{f}(p) = f(x)$ if $p= \pi(x)$. This map is well defined, since $f(x) = f(y)$ whenever $\pi(x) = \pi(y)$. 

\begin{figure}[!ht]
    \centering
    \includegraphics[width=0.5\textwidth]{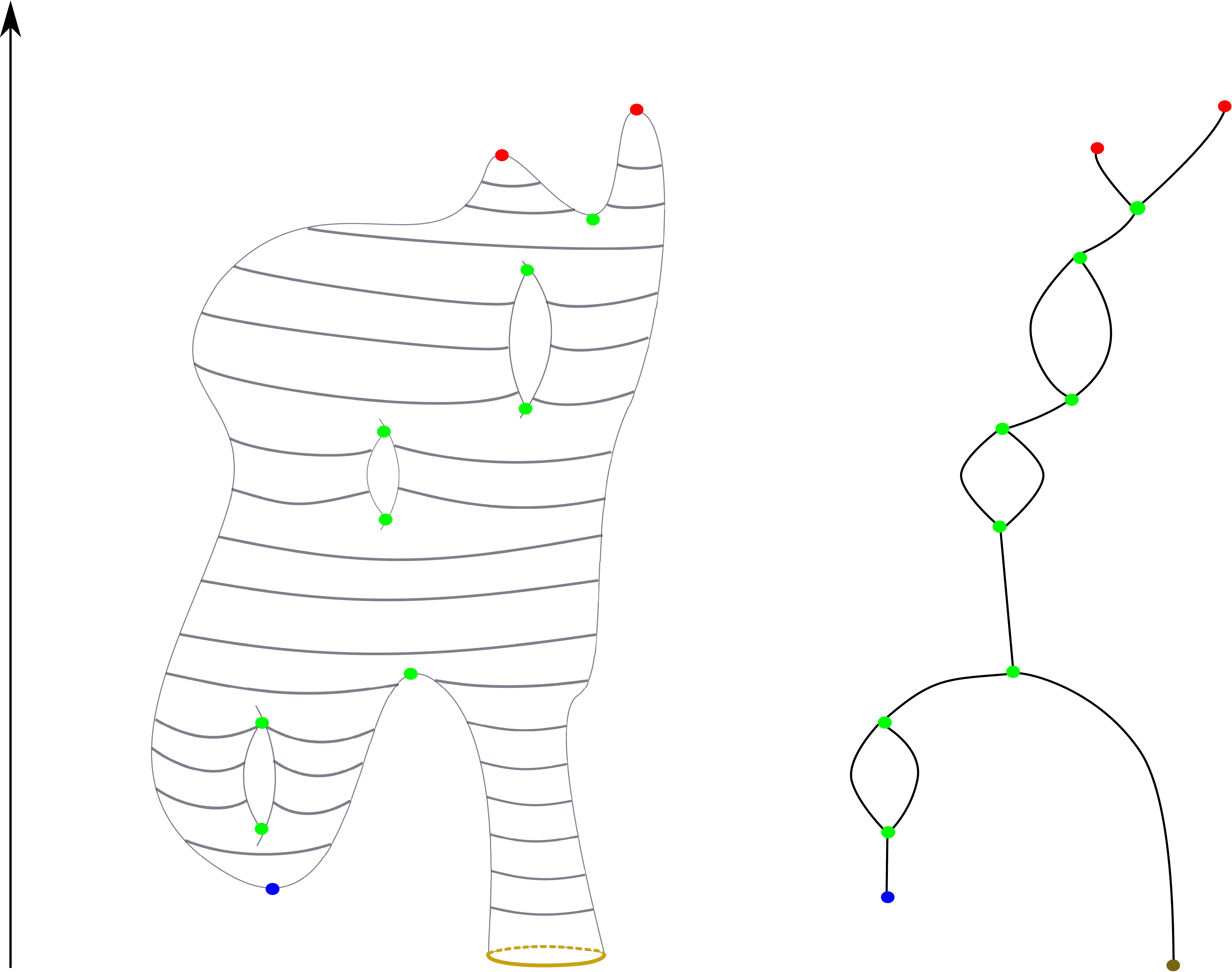}
    \put(-240,165){$f$}
    \caption{An example of a Reeb graph (right) of a scalar function (left) defined on a surface with a boundary.}
    \label{Reebexample}
\end{figure}

When $M$ is a manifold, and $f$ is Morse, then $R(M,f)$ exhibits certain additional properties. For instance, in this case, every vertex of $R(X,f)$ arises from a critical point of $f$ or a boundary component. Furthermore,  every maximum or minimum of $f$ gives rise to a degree $1$-node of $R(M,f)$. Saddle points for a Morse function $f$ defined on a $2$-manifold have degree $3$-node. This is not guaranteed if the scalar function is not Morse. See Figure~\ref{Reebexample} for an example of a Reeb graph.

\section{Reeb Graphs on Triangulated Surfaces}

\subsection{Morse Functions in the Piece-Wise Linear Setting}

\begin{figure}[!ht]
    \centering
    \includegraphics[scale=.15]{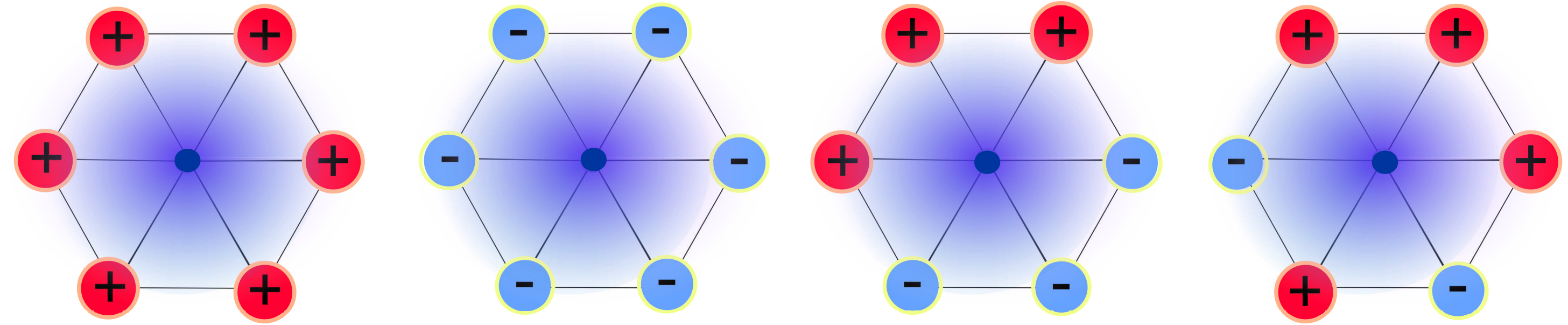}
    \caption{The types of vertices on a triangulated mesh. From left to right: minimum, maximum, regular vertex, saddle.}
    \label{classification}
\end{figure}

In this paper, we will be working with triangulated surfaces. We will denote the set of vertices, edges, and faces of a triangulated mesh $M$ by $V(M)$, $E(M)$, and $F(M)$, respectively. The extension of Morse theory to triangulated manifolds was given by Banchoff ~\cite{critical1967}. Let $M$ be a triangulated $2$-manifold and let $f :M\longrightarrow [a,b]\subset \mathbb{R}$ be a piece-wise linear function on $M$. We will use the abbreviation PL for "piece-wise linear". The star of $v$, denoted by $star(v)$, is the set of simplices that intersect with the vertex $v$. The \textit{closure} $\overline{star(v)}$ of $star(v)$ is the smallest simplicial subcomplex of $M$ that contains $star(v)$. The \textit{link} $Lk(v)$ consists of the subcomplex of $M$ of simplices belonging to $\overline{star(v)}$ but not to $star(v)$. The \textit{upper link} of $v$ is defined to be the set:
\begin{equation*}
Lk^+(v)=\{ u \in Lk(v):   f(u)>f(v)\}\cup\{ [u,v]\in \overline{star(v)}: f(u)>f(v) \},
\end{equation*}
the lower link is defined similarly by:
\begin{equation*}
Lk^{-}(v)=\{u \in Lk(v):  f(u)<f(v)\}\cup\{ [u,v]\in \overline{star(v)}: f(u)<f(v) \},
\end{equation*}
and \textit{mixed link} 
\begin{equation*}
Lk^{\pm}(v)=\{ [u_1,u_2] \in Lk(v) : f(u_1)< f(v) < f(u_2)\}.
\end{equation*}
Using the link definitions, we classify vertices of $M$ as follows. A vertex $v$ in $M$ is \textit{PL regular} if the cardinality $|Lk^{\pm}(v)|$ of $Lk^{\pm}(v)$ is equal to $2$. If $|Lk^+(v)|=0$, then $v$ is a \textit{PL maximum} vertex with index $1$, and if $|Lk^-(v)|=0$, then $v$ is a \textit{PL minimum} with index $0$. If $|Lk^{\pm}(v)|=2+2m$, then $v$ is a PL saddle with index $1$ and multiplicity $m\geq 1$. See Figure~\ref{classification}. A PL function $f:V\longrightarrow [a,b] \subset \mathbb{R}$ is a \textit{PL Morse function} if each vertex is either PL regular or PL simple, and the function values of the vertices are distinct. Similar to the smooth case, when $\partial M \neq \emptyset $, we assume further that $f$ satisfies: (1) $f^{-1}(\partial [a,b]) = \partial M$ and (2) there are no critical points on $\partial M$.

\subsection{Reeb Graphs of General Simplicial Complexes}
Reeb graph can be defined naturally on arbitrary simplicial complexes. Let $K$ be a simplicial complex and  $f:V(K) \longrightarrow \mathbb{R}$ be a map defined on the vertices of $K$. The map $f$ can be extended linearly to all simplices of $K$ to PL function, which we will denote also denote by $f$. Using this function, the Reeb graph of $(K,f)$ can be defined as before.

\section{Reeb Graph Sequential Algorithm on Triangulated Surfaces}

In this section, we assume that we are given a PL function $f:M\longrightarrow \mathbb{R}$ defined on a triangulated surface $M$ possibly with a non-empty boundary $\partial M$. This includes the case when $f$ is a Morse function. Then we discuss the degenerate case when $f$ has non-simple saddles.

The algorithm above relies on Morse theory to find a finite set of paths traced concurrently inside parts of the manifold where the topology of the manifold with respect to a given scalar function does not change. The sequential algorithm that we present here is similar to the Reeb graph algorithm given in~\cite{doraiswamy2008efficient}, where tracing paths inside cylinders were used to construct the Reeb graph. We provide the sequential version of the algorithm because it has some key differences from the algorithm given~\cite{doraiswamy2008efficient} that will be utilized later in our parallel algorithm.

The main idea of the algorithm is the construction of a sub-simplicial complex $X$ of $M$, such that the Reeb graph $R(X,f|_X)$ of $X$ with respect to $f|_X$, the restriction of $f$ on $X$, is identical to the Reeb graph $R(M,f)$. The constructed simplicial complex $X$ does not only provide us with a Reeb graph of $(M,f)$ but also implies immediately an algorithm to compute the map $F:R(M,f)\longrightarrow M$ that allows us to extract the manifold data given the corresponding Reeb graph points. The main two ingredients of the algorithm are the \textit{critical sets} and the \textit{ascending paths}. We introduce these two concepts next.

\subsection{Critical Sets and Ascending Paths}
\label{critical sets}
We start by giving the definition of critical sets. Then, we provide the definition of ascending paths.

\begin{figure}[!t]
    \centering
    \includegraphics[width=0.4\textwidth]{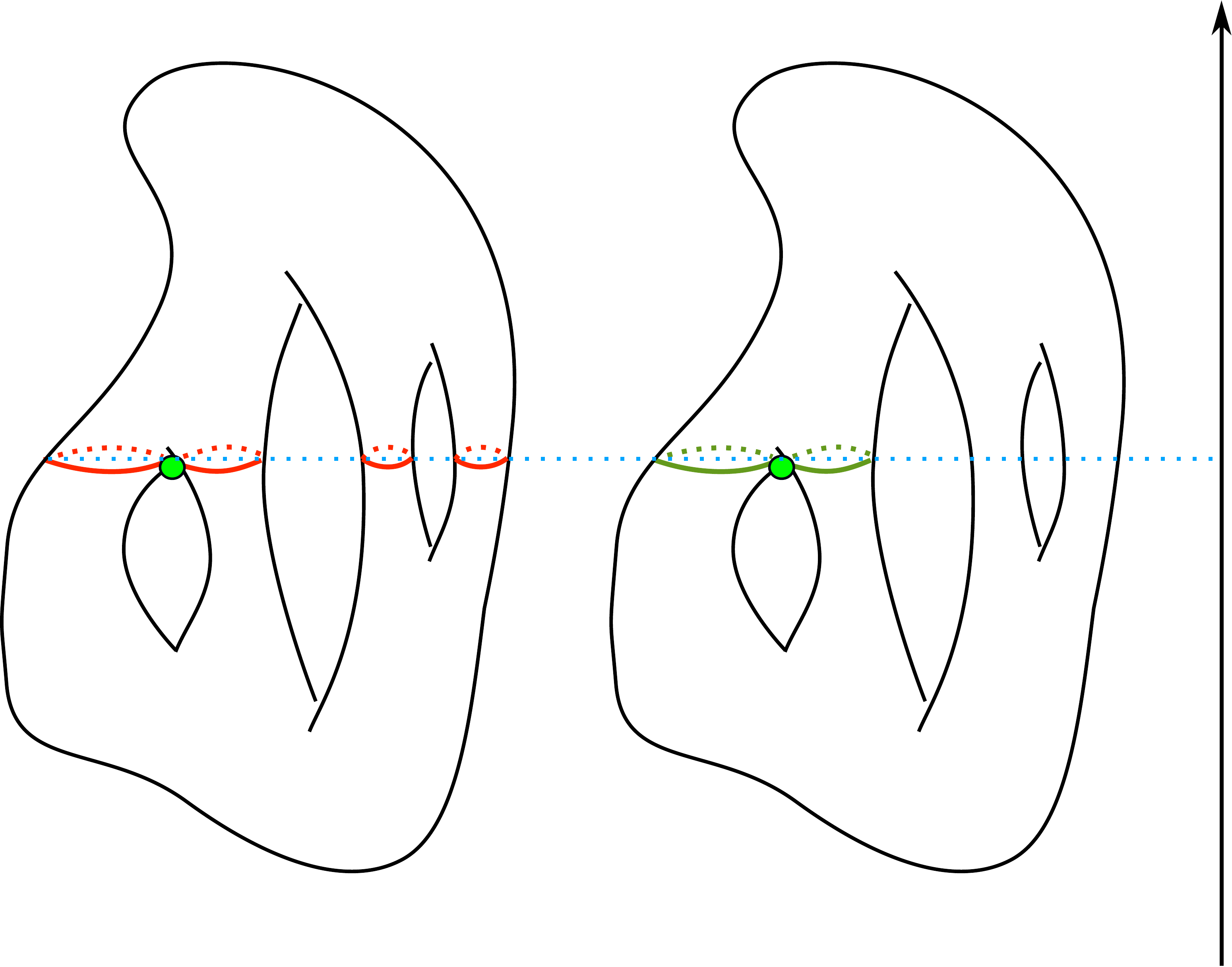}
    \put(-165,84){$p$}
    \put(-70,84){$p$}
    \put(0,80){$t_p$}
    \caption{For $t_p$ a critical value, on the left, the set $f^{-1}(t_p)$ consists of a singular set $C_p$ and two of simple closed curves. On the right, only the critical set $C_p$ is shown, which is the connected component of $f^{-1}(t_p)$ that contains the critical point $p$.}
    \label{Critical set}
\end{figure}

\para{Critical Sets}
Let $p$ be a saddle point of $f$, and let $t_p$ be its corresponding critical value. Consider the connected components of the set $f^{-1}(t_p)$. The connected components of $f^{-1}(t_p)$ consist of a collection of simple closed curves embedded in $M$, as well as a single component, which contains a singularity. This \textit{singular set} consists of multiple circles that intersect at the critical point $p$. We will denote this singular set by $C_{p}$. See Figure~\ref{Critical set} for an example. Note that for critical value $t$ $f^{-1}(t_p)$ might consist merely of the critical set $C_{p}$ (with no other simple closed curves).

Choose $\epsilon >0$ small enough such that the interval $[t_p-\epsilon,t_p+\epsilon]$ has only the critical value $t_p$. As we move from $t_p$ to $t_p-\epsilon$, the singular set $C_p$ becomes a non-singular one consisting of a disjoint union of simple closed curves $A_1,...A_n$, for $n\geq 1$. By convention, we will consider the sets $A_1,...,A_n$ to be the \textit{connected components of the singular set} $C_p$, and we will refer to them as such for the rest of the paper. We talk more about the components of a critical set and show exactly how to determine them in the piece-wise linear setting in Section~\ref{critical sets details}. The following Lemma asserts that the number of connected components of $C_p$ of for a simple saddle point of a Morse function defined on the surface is either $1$ or $2$.

\begin{figure}[!t]
    \centering
    \includegraphics[width=0.4\textwidth]{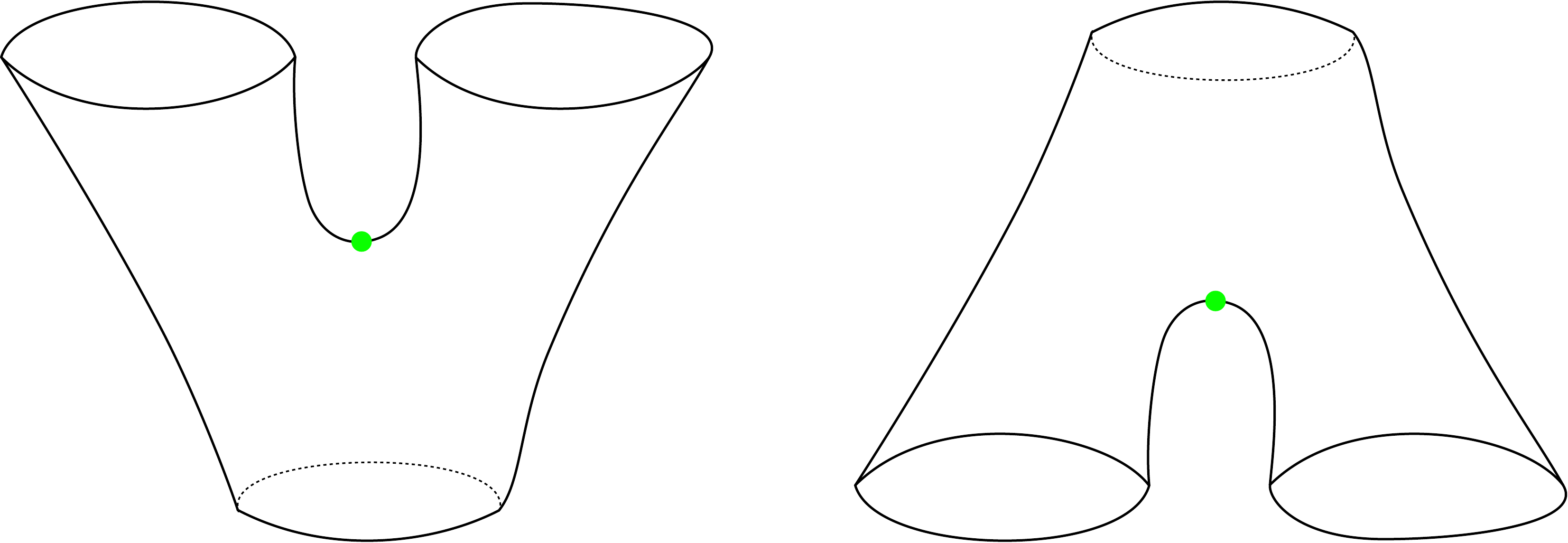}
    \put(-150,41){$p$}
    \put(-50,34){$p$}
    \caption{A split saddle on the left and a merge saddle on the right.}
    \label{pants}
\end{figure}

\begin{lemma}
\label{lemma}
    Let $M$ be a compact connected orientable surface with more than one boundary component. Let $M\longrightarrow [a,b]$ be a Morse function on $M$  such that $f^{-1}(\partial ([a,b]))= \partial M $. If $f$ has a unique saddle point, then $M$ is homeomorphic to a pair of pants. See Figure~\ref{pants}.
\end{lemma}

For a regular value $t$, we will denote the number of simple closed curves of $f^{-1}(t)$ by $|f^{-1}(t)|$. Lemma~\ref{lemma} implies that for a sufficiently small enough $\epsilon$ and for any saddle point $p$ on a Morse function $f$, one has $|f^{-1}(t_p+\epsilon)|-|f^{-1}(t_p-\epsilon)|=\pm 1$. In other words, as we are passing through a saddle point $p$ two circles merge or split. Figure~\ref{pants} shows two types of saddles: \text{a split saddle and a merge saddle}. A saddle point $p$ is a split saddle if $|f^{-1}(t_p+\epsilon)|-|f^{-1}(t_p-\epsilon)|= 1$, and it is a merge saddle if $|f^{-1}(t_p+\epsilon)|-|f^{-1}(t_p-\epsilon)|= -1$.

The sequential version of the algorithm relies on Lemma~\ref{lemma} to handle the case when the function $f$ is Morse and has only simple saddle points. The case when $f$ has saddle points with higher multiplicities will be handled in Section~\ref{degenerate}.

\begin{remark}
    If $p$ is a maximum or minimum point, then by definition, $C_p$ will be the set that consists of the point $p$ itself.
\end{remark}

\para{Ascending Paths} The second main ingredient of the sequential Reeb graph algorithm is a collection of curves that we trace inside the manifold $M$ using the function values. More precisely, an \textit{ascending path} from a non-maximum vertex $v_0$, denoted by $apath(v_0)$, is defined to be a finite sequence of consecutive edges $\{[v_0,v_1],...,[v_{k-1},v_k]\}$ on $M$, such that $[v_i,v_{i+1}]$ is an edge on $M$ for $0\leq i \leq k-1$, $f(v_{i+1})>f(v_{i})$, and $v_k$ is a maximum, a boundary or a saddle vertex.

\subsection{Outline of the Sequential Algorithm}
\label{alg}

We present now the outline of the sequential Reeb graph algorithm. We assume that we are given a triangulated PL Morse function $f:M \longrightarrow [a,b] \subset \mathbb{R}$ defined on triangulated mesh $M$ without boundary. The case when the function $f$ is not Morse, or when the $M$ has a boundary, will be discussed in later sections. 

The sequential Reeb graph algorithm is given in the following steps: 

\begin{enumerate}
    \item We start by sorting the critical points of $f$ by their critical values. Let $CP$ be the set of the sorted critical points of $f$ in an ascending order. 

    \item  For each critical point of the function $f$, we define a node in the Reeb graph $R(M,f)$. In other words, the node set of the graph $R(M,f)$ corresponds precisely to the set of critical points of the function $f$ defined on $M$.

    \item For each critical point $v$ in $CP$, we compute the critical set $C_v$.

    \item For each saddle or a minimum vertex $v$ in $CP$, we associate one or two ascending paths on the mesh: one ascending path if $v$ is a minimum or a merge saddle, and two paths if $v$ is a split saddle. For each ascending path, we march with until this path intersects with the first critical $C_w$ set with a higher critical value than of $t_v$. At this point, we insert an edge for the Reeb graph $R(M,f)$ between the vertex $v$ and the vertex $w$.
\end{enumerate}

\begin{figure}[!ht]
    \centering
    
    \includegraphics[trim=0 1675pt 0 0, clip, width=0.4\textwidth]{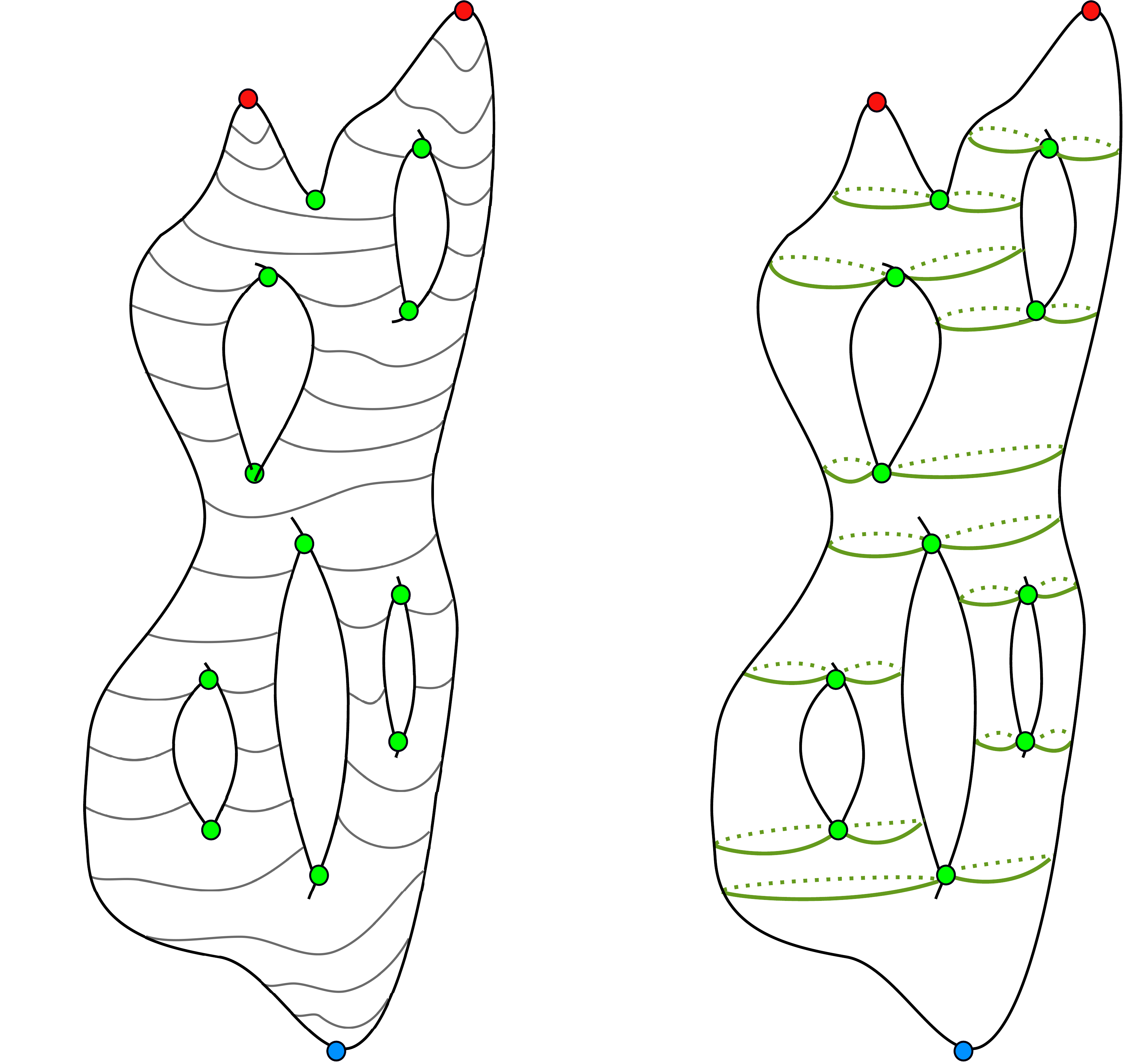}\hspace{40pt}
    \includegraphics[trim=0 0 0 1665pt, clip, width=0.4\textwidth]{figures/REEB_GRAPH_ALG_1-eps-converted-to.pdf}
    \put(-410,145){$(1)$}
    \put(-310,145){$(2)$}
    \put(-200,145){$(3)$}
    \put(-80,145){$(4)$}
    
    \caption{Summary of the sequential algorithm. (1) The input of the algorithm is a manifold $M$ with a scalar function $f$. (2) Computing the critical sets. (3) For each saddle point or minimum, we compute the ascending paths (4) creating the edges of the Reeb graph by the information encoded in the start vertex of the ascending vertex and its termination critical set.}
    \label{Reebexamplealg}
\end{figure}

Figure~\ref{Reebexamplealg} illustrates the steps of the algorithm. It remains to describe two aspects of the previous algorithm: the construction of the critical sets mentioned in step (3) and the construction of the ascending paths mentioned in step (4).

\subsubsection{Construction of the Critical Sets}  
\label{critical sets details}
We now describe how to compute the critical set of a critical point in the piece-wise setting. As before, we assume that $f:M\longrightarrow [a,b] \subset \mathbb{R}$ a piece-wise linear function is defined on a triangulated surface $M$, and it takes distinct values on the vertices of $M$. This assumption will guarantee that for a given value $t$, the level curve $f^{-1}(t)$ intersects with at most one vertex of $M$.

Now let $t \in \mathbb{R}$. The \textit{cross simplices} $CR_f(t)$ of the value $t $ is the union of all simplices of $M$, which intersect with the level curve $f^{-1}(t)$. This is the set of vertices, edges, and faces in $M$, which intersect with the level curve $f^{-1}(t)$. We define $\overline{CR_f(t)}$ to be the closure of the smallest subcomplex of $M$ which contains $CR_f(t)$. If $f^{-1}(t)=v$ for a vertex $v$ in $V(M)$, then we define the subcomplex $\overline{ CR_f(t)}$ as above but we also add to it the simplices of   $\overline{star(v)}$.

\begin{figure}[!ht]
    \centering
    \includegraphics[width=0.5\textwidth]{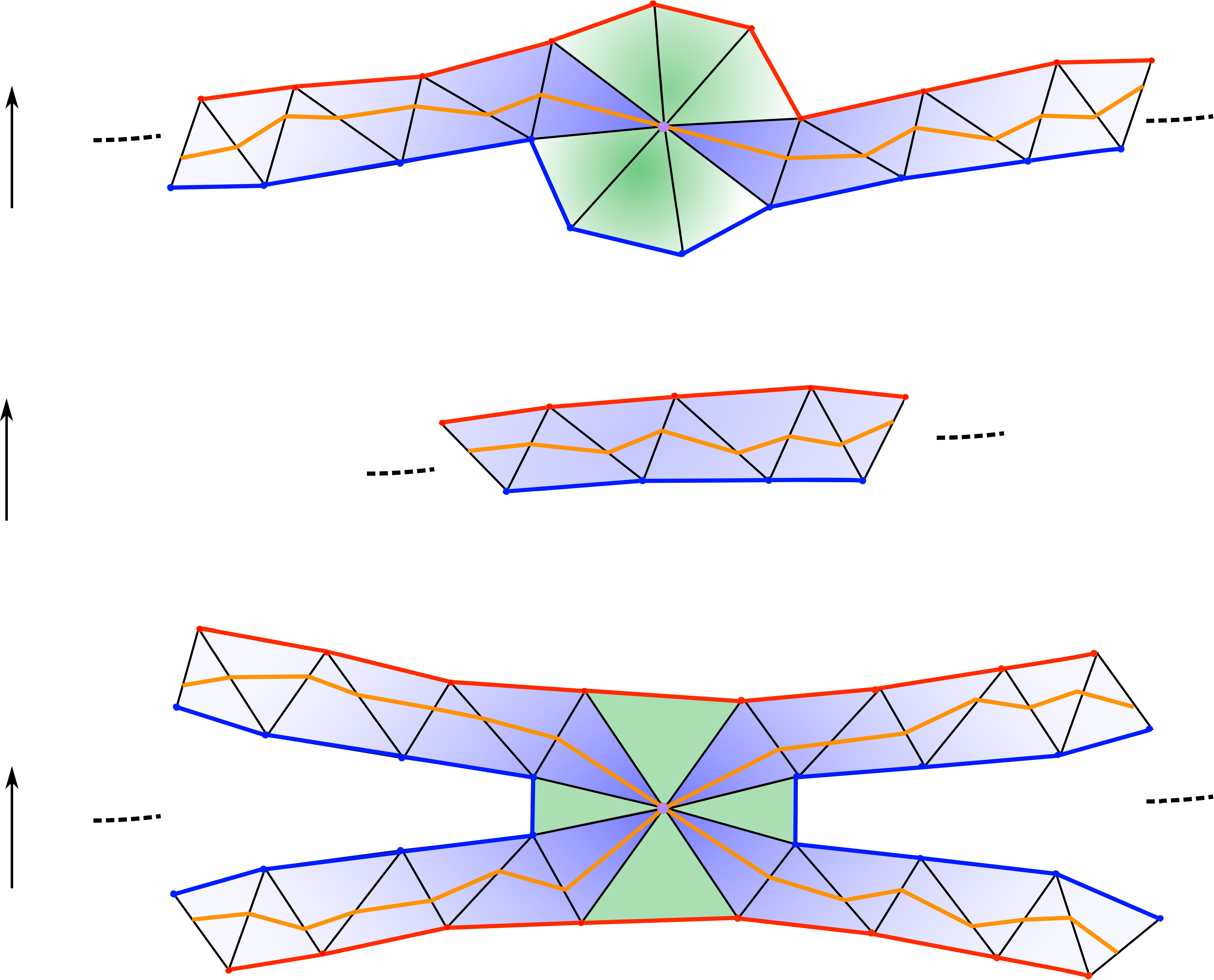}

    \caption{The set of vertices, edges, and faces in the mesh $M$, which $f^{-1}(t)$ intersects. The top two figures depict cases where $t$ is a regular value. The top figure shows an example of when there exists a vertex $v$ in $M$, such that $v \in f^{-1}(t)$. The second figure shows the case where no such a vertex exists. In other words, $f^{-1} (t)$ intersects only edges and faces of $M$. The third figure is an example of the local neighborhood, a critical point of $f$ or when $f^{-1}(t)$ contains a critical point. The purple simplices represent the set $ CR_f(t)$. The union of the green and purple simplices represents the set $\overline{ CR_f(t) }$.  The blue edges and nodes are the edges in the complexes shown in the figure and nodes in $\overline{CR_f(t)}$, which have $f$-values less than or equal to $t$.  The red edges and nodes are the edges in the complexes shown in the figure and nodes in $\overline{CR_f(t)}$, which have $f$-values higher than or equal to $t$.} 
    \label{levelset1}
\end{figure}

When $t$ is the maximal or the minimal value, then $CR_f(t)$ consists of a single vertex $v_t$. In this case, $\overline{ CR_f(t)}$ is simply $\overline{star(v_t)}$. When $t$ is a regular value, then $\overline{CR_f(t)}$ is a disjoint union of  topological cylinders, that is, $\overline{CR_f(t)}$ appears as a "thickened" band around the curve $f^{-1}(t).$ Note that when the value $t$ corresponding to a vertex $v$ in $V(M)$ with $f(v)=t$ for some $v$ in $M$, then the set of simplices in the intersection $CR_f(t) \cap f^{-1}(t)$ is simply the vertex $\{v\}$. When $t$ is the critical value of a saddle point, then the curve $f^{-1}(t)$ consists of a finite collection of simple close curves that meet at the saddle point, and the set $\overline{ CR_f(t)}$ can be seen as the thickened band of these curves. See Figure~\ref{levelset1}. 

The \textit{lower level} of the cross subsimplex of a value $t$, denoted by $L(\overline{ CR_f(t) } )$, is the set of vertices and edges in $\overline{ CR_f(t) }$, which have values less than or equal to $t$. Similarly, the \textit{higher level} of the cross subsimplex of a value $t$, denoted by $H(\overline{ CR_f(t) } )$, is the set of vertices and edges in $\overline{ CR_f(t) } $, which have values higher than or equal to $t$. See Figure~\ref{levelset1}.

Recall the notion of the connected component of a critical set from Section~\ref{critical sets}. Specifically, for a critical vertex $v$ with a critical value $t_v$, we talked about the connected components of the critical set $ C_{v}$. Using the definitions introduced in this section, we can compute the connected components of the critical set $ C_{v}$ in the piece-wise linear setting by considering the connected components of $L(\overline{ CR_f(t_v)})$.

\subsubsection{Construction of the Ascending Paths}
\label{ascending}
The ascending paths from a critical point $v_0$ are specified as follows:

\begin{enumerate}

    \item If $v_0$ is a merge saddle or a minimum, then we initiate a single ascending path $apath(v_0)$ specified as follows. Let $v_1$  be a vertex in $Lk^+(v_0)$ such that $f(v_1)>f(v_0)$. At the $k^{th}$ iteration, $apath(v_0)$ consists of $\{[v_0,v_1],...[v_{k-1},v_k]\}$ with $f(v_{i+1})>f(v_{i})$ for $0\leq i \leq k-1$.

    \item If $v_0$ is a split saddle, then we start two ascending paths $P_1$ and $P_2$ originating from the point $p$ specified as follows. Divide the set $Lk^+(p)$ into two disconnected components $A$ and $B$. Choose the vertex $v_A$ in $A$, such that $f(v_A)> f(v)$ for all $v \in A$, and choose the vertex $v_B$ in a similar manner. At the $k^{th}$ iteration, $P_1$ consists of $\{[v_0,v_A],...[v_{k-1},v_k]\}$ with $f(v_i)>f(v_{i-1})$ for $0\leq i \leq k-1$ (here we assume $v_1=v_A$). The path $P_2$ is constructed similarly.
\end{enumerate}

\subsubsection{Termination of an Ascending Path}

The condition at which we terminate the ascending paths we initiated in step $(4)$ is specified as follows.  Assume that we initiated an ascending path from a critical vertex $v$. Let $w$ be the critical vertex with the critical value $t_w$ right after the critical value $t_v$ of $v$. Assume that at the $k^{th}$ iteration, an ascending path starting from the vertex $v$ is $\{[v,v_1],...[v_{k-1},v_k]\}$. We continue this iteration until we arrive at an edge $E_n=[v_{n-1},v_n]$ with $f(v_n)\geq t_w$ and $f(v_{n-1} )<  t_w$. At this point, we check the condition $C_{w}\cap E_{n} \neq \emptyset$. If this condition is satisfied, then we insert an edge for the Reeb graph $R(M,f)$ between the vertex $v$ and the vertex $w$. If  $C_{w}\cap E_{n} = \emptyset$, then we keep marching until an edge in the ascending path meets a critical point $w$ that satisfy these two conditions. Note that $w$ can be either a saddle or a maximum vertex.

The check of intersection between the edge $E_n$ in an ascending path and a critical set $C_{w}$, mentioned in step (4), can be done by checking if $E_n$ belongs with $CR_f(w)$.

\begin{remark}
    \label{potential}
    It is important to notice how an ascending path corresponds to an edge in the Reeb graph. The ascending path starts at a critical point $p$ with a critical value $t_p$ and terminates at a critical set that corresponds to a critical point $q$ with a critical value $t_q$ with $t_q>t_p$. More precisely, an ascending path starts at one of the connected components of the upper link of a critical point $p$ and ends at one of the connected components of the critical sets $C_{q}$. If two ascending paths start at two different connected components of the upper link of $p$ but still end up in the \textit{same connected component} of $C_{q}$, then these two paths correspond to the exact same edge in the Reeb graph. Therefore only one of these ascending paths corresponds to an edge in the final Reeb graph. For this reason, we say that each ascending path starting from a connected component of the upper star of a vertex gives rise to a \textit{potential} edge in the Reeb graph. We provide more details on this point in Section~\ref{degenerate}.  
\end{remark}

\subsection{Surfaces with Boundaries}
In the case when the surface $M$ has a boundary, we modify the previous algorithm as follows. In this case, $f^{-1}(\partial I)= f^{-1}({a}) \cup f^{-1}({b})$ is not empty and consists of a finite collection of simply closed curves. We treat each connected component of $f^{-1}(a)$ as a minimal point, and we treat the boundary $f^{-1}(b)$ as a maximum point. More precisely, the following modifications are added to the previous algorithm from Section~\ref{alg}.

\begin{itemize}
    \item In step $(2)$, each connected component in $f^{-1}(\partial I)$ is considered a vertex in the Reeb graph vertex set.

    \item In step $(3)$, for each boundary component in $f^{-1}(a)$, we pick an arbitrary vertex on the boundary and initiate an ascending path starting from that vertex.

    \item In step $(4)$, if an ascending path starting at a vertex $v$ reaches a boundary vertex $w$ in one of the connected components, say $Bndry_w$, of $f^{-1}(b)$, then we insert an edge in the Reeb graph $R(M,f)$ between the vertex $v$ and the vertex in $R(M,f)$ that corresponds to $Bndry_w$.
\end{itemize}

We denote the subcomplex obtained from $M$ using the previous algorithm by $X_{M,f}$. In other words, $X_{M,f}$ consists of the critical sets $C_p$ for all critical points $p$, as well as the ascending paths we initiated at the saddle, minimum, or boundary vertices. When $M$ and $f$ are clear from the context, we will denote to $X_{M,f}$ simply by $X$.

\subsection{Correctness of the Sequential Algorithm}

For a function $f:M\longrightarrow [a,b]$ and $c \in \mathbb{R}$  define: 
$$M_c:=\{ x\in M |f(x) \leq c \}.$$
Note that we allow $M_c$ to be empty. Moreover, we define:
$$M_{[c,d]}:=\{ x\in M |c \leq  f(x) \leq d   \}. $$

The correctness of our algorithm relies on the following two facts:

\begin{enumerate}
    \item The only topological changes to the level sets of $f$ occur when as pass a critical point. This is stated formally in Theorem~\ref{same topology}.
    
    \item The structure of the manifold around a critical point is completely determined by the index of that critical point. We give this in Theorem~\ref{main thm}. 
\end{enumerate}
The proof of Theorems~\ref{same topology} and~\ref{main thm} can be found in~\cite{milnor1963morse}.

\begin{theorem}
    \label{same topology}
    Let $f : M \longrightarrow [a^{\prime},b^{\prime}]$ be a smooth function on a smooth surface M. For two reals $a,b$ with $a^{\prime}< a <  b < b^{\prime}  $, if $f$ has no critical values in the interval $[a, b]$, then the surfaces $M_a$ and $M_b$ are homeomorphic.
\end{theorem}

The algorithm above relies on Theorem~\ref{same topology}. Namely, as we trace an ascending path, we assume that no topological change occurs until we reach the next critical point. The ascending path may not terminate at the next critical point, provided the part of the manifold in which this path is traveling in has not changed its topology. This will become more evident after we provide the next theorem, which shows the exact structure of $M_{[c,d]}$ around a critical point.

\begin{theorem}
    \label{main thm}
    Let $M$ be a compact connected surface, possibly with a boundary, and let $f:M\longrightarrow [a,b]$ be a Morse function on $M$. Let $p$ be a critical point, and let $t$ be its corresponding critical value. Let $\epsilon >0$ be small enough so that $I_\epsilon:=[\epsilon-t,\epsilon +t]$ has only the critical value $t$.  

    \begin{enumerate}
        \item If $index(p)$ is equal to 0 or 2, then $M_{I_\epsilon}$ is homeomorphic to a disjoint union of a disk and a finite number of topological cylinders (that is a genus zero surface with two boundary components).

        \item If $index(p)=1$, then $M_{I_\epsilon}$ is homeomorphic to a pair of pants and a finite number of topological cylinders. 

    \end{enumerate}
\end{theorem}

\begin{figure}[!ht]
    \centering   
    \includegraphics[angle=90,origin=c,width=0.4\textwidth]{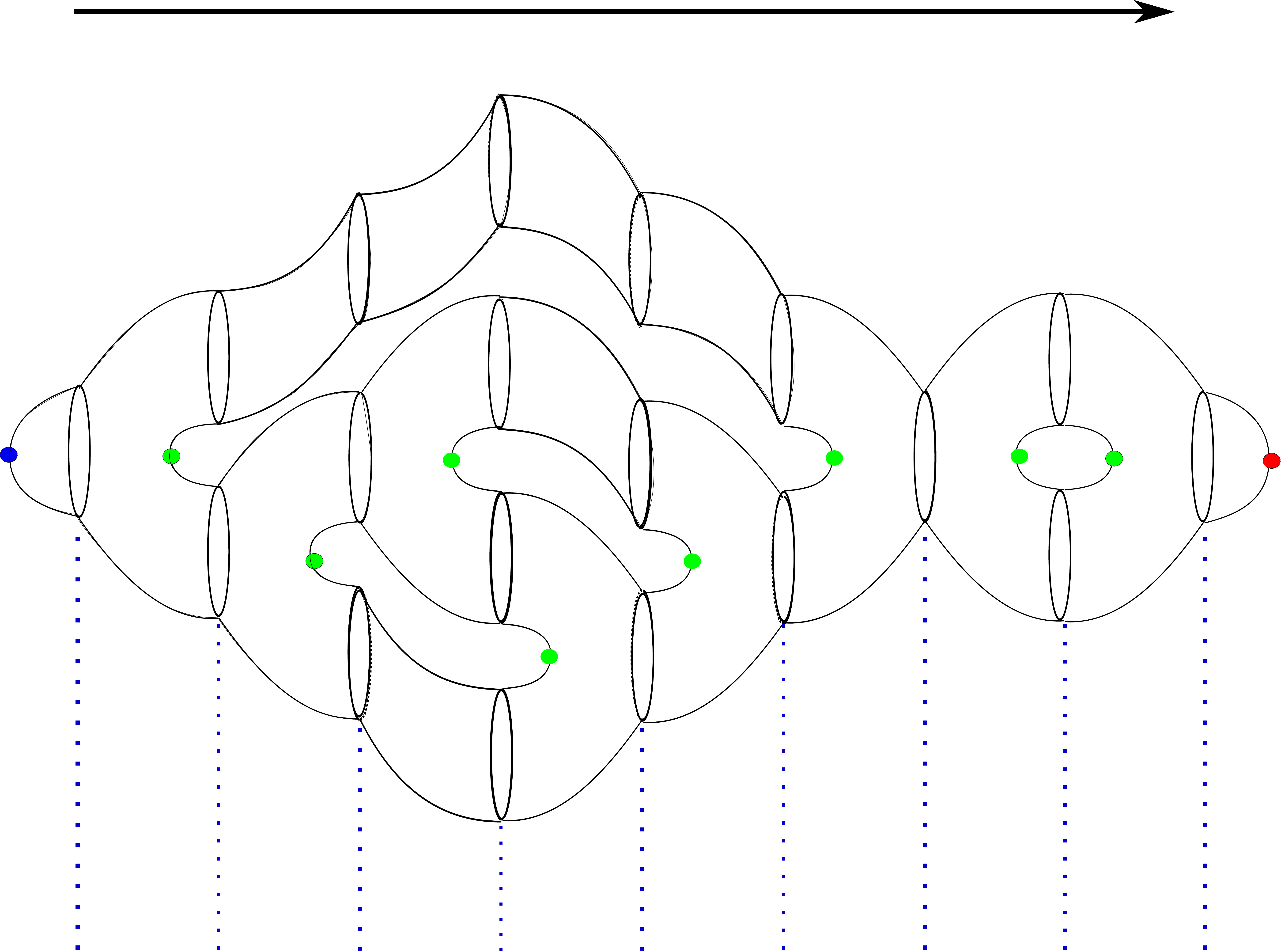}

    \caption{Given a Morse function $f$ on a surface $M$, we can slice $M$ so that around each critical value $t$ the submanifold $M_{[t-\epsilon,t+\epsilon]}$  is a disjoint union of simple building blocks: pair of pants, topological cylinders, and topological disks.}
    \label{correctness}
\end{figure}

The previous two theorems show that for a given Morse $f$ function on $M$, we can arrange $M$ so that at each critical point, only a single topological event occurs, and this topological event occurs around the critical point. Moreover, we know exactly what topological event occurs by considering the index of the critical point. This is illustrated in Figure~\ref{correctness}.

\begin{figure}[!ht]
    \centering
    \includegraphics[width=0.6\textwidth]{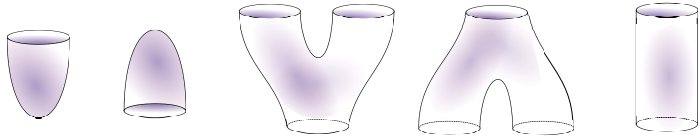}

    \caption{The building blocks of a surface. Given a scalar function $f$ defined on a surface $M$, Theorem~\ref{main thm} asserts that we can decompose the surface into the building block pieces appear in above. These pieces are genus zero surfaces with a single boundary component (disk), genus zero surfaces with 3 boundary components (pair of pants), and a genus zero surface with 2 boundary components (cylinder).}
    \label{building blocks}
\end{figure}

\begin{figure}[!ht]
    \centering
    \includegraphics[width=0.6\textwidth]{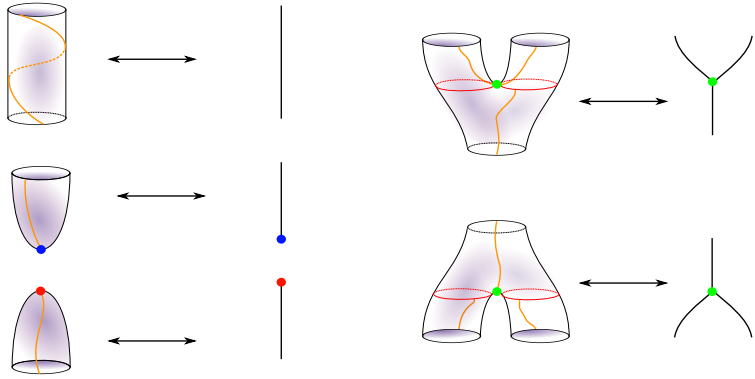}

    \caption{The restriction of the Reeb graph on the building blocks of a surface along with the part of the ascending paths in that part of the surface. Around each critical point, the structure of the ascending curves (orange) is identical to the structure of the Reeb graph. More precisely, the quotient space of the restriction of the function on the ascending curves the critical sets around a critical point (these are the curves highlighted by orange and red in the figure) is identical to the quotient space of the manifold locally.}
    \label{ascending paths}
\end{figure}

This shows that any 2-manifold can be built from the building blocks in Figure~\ref{building blocks}. Moreover, Theorem~\ref{main thm} shows that the restriction of $f$ on $M_{I_\epsilon}$ has the shapes given in Figure~\ref{ascending paths}. In other words, this gives us the structure of the Reeb graph around an interval $I_{\epsilon}$ that contains a single critical value.

Moreover, subcomplex $X_{M,f}$ constructed in the algorithm has the same Reeb graph structure of that of $M$ around the critical points. See Figure~\ref{ascending paths}.

Now let $e=(p,q)$ be an edge in a Reeb graph $R(M,f)$ connecting between the two nodes $p$ and $q$, which correspond to critical points of $f$. For each such edge, we can find a preimage arc $E$ in $M$ that is mapped to $R(M,f)$ via the map $\pi: M\longrightarrow R(M,f)$. The preimages of the value $p$ and $q$ under $\pi$ are the critical set $C_p$ and $C_q$ in $M$, respectively. Hence each arc $E$ must start at a point $A$ in $C_p$ and end at a point in $B$ in $C_q$. Each such an arc on $M$ characterizes the edge $e$.

An ascending path constructed in the sequential algorithm essentially traces an arc in the way described above. Namely, for a critical point $p$, with a critical value $f(p)$, an ascending path created at a point $p$ will terminate at a critical point $q$ with $f(q)>f(p)$. This termination occurs when we pass through the critical set of the point $q$.

\subsection{Dealing With Degenerate Cases}
\label{degenerate}
It is possible in practice to obtain a scalar function with saddle points that have multiplicity $m \geq  2$. The algorithm that we present in Section~\ref{alg} can be extended to handle such cases. We need to make the following modifications.

\begin{itemize}
    \item In step (3), we calculate the connected components of the critical set $C_p$. Here it is not enough to calculate the critical set $C_p$. We also need to compute the connected components of this set. This was explained in Section~\ref{critical sets details}.
    
    \item When the multiplicity $m$ of a split saddle $p$ is greater than or equal to $2$, we create an ascending path for each connected component in $Lk^+(p)$. 
\end{itemize}

In the non-degenerate case, every ascending path corresponds to an edge in the Reeb graph. This is not the case anymore in the degenerate case. We describe next the sequential algorithm of the Reeb graph when $f$ has degenerate critical points.

\begin{enumerate}
    \item Sort the critical points of $f$ an in ascending order. Let $CP=\{v_1,...,v_n\}$ be this set. This represents the vertex set of the Reeb graph, as we did before.

    \item For each $v$ in $CP$, compute the connected components of $C_v$ and include all these components in a single container $\mathcal{S}$. We will denote by $B^j_i$ to be the connected component $j$ of the critical set $C_{v_i}$. In this way, we index all elements in $\mathcal{S}$.

    \item Declare each set $B^j_i$ of $\mathcal{S}$ as not visited. 

    \item For each critical point $v_i$ in $CP$ and for each component in $Lk^{+}(v_i)$, we initiate an ascending path $P$ as described in Section~\ref{ascending}. For each such path, we determine the connected component $B^l_k$ in $\mathcal{S}$, which the path terminates as described in Section~\ref{critical sets details}. We have two cases:

    \begin{enumerate}
        \item If the connected component $B^l_k$ is not visited, then we insert an edge between $v_i$ and $v_k$ and mark the component $B^l_k$  as visited.

        \item If the connected component $B^l_k$ is visited, then we do not make any changes to the Reeb graph and terminate the current ascending path. In this case, the ascending path corresponds to an edge that already exists in the Reeb graph. See Remark~\ref{potential}.
    \end{enumerate}

\end{enumerate}

Note that the above procedure can be used to determine the edges originating from a simple saddle. Namely, we do not need to check the type of the simple saddle point (merge or split) in step (3) of the algorithm given in Section~\ref{alg}, and for any saddle point, we use the above procedure instead.

\section{Parallelization of the Algorithm}

In this section, we give the details of our strategy to compute the Reeb graph in parallel. We describe the three stages of the parallel algorithm as follows.

\begin{enumerate}
    \item \textbf{The Partition Stage}. In this stage, we partition the manifold $M$ into submanifolds such that the vertices counts of each submanifold  are approximately equal to each other.
    
    \item \textbf{Computing the Reeb Graph for the Submanifold Stage}. Computing the Reeb graphs for each submanifold obtained from stage one concurrently.
    
    \item \textbf{The Gluing Stage}. In this step, we glue the Reeb graphs obtained from stage 2.
\end{enumerate}

\subsection{The Partition Stage}
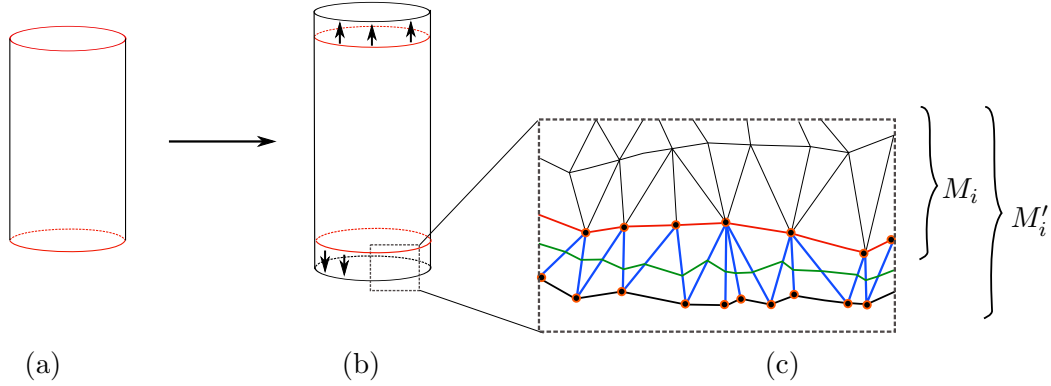
\begin{figure}[!ht]
    \centering
    \includegraphics[width=0.8\textwidth]{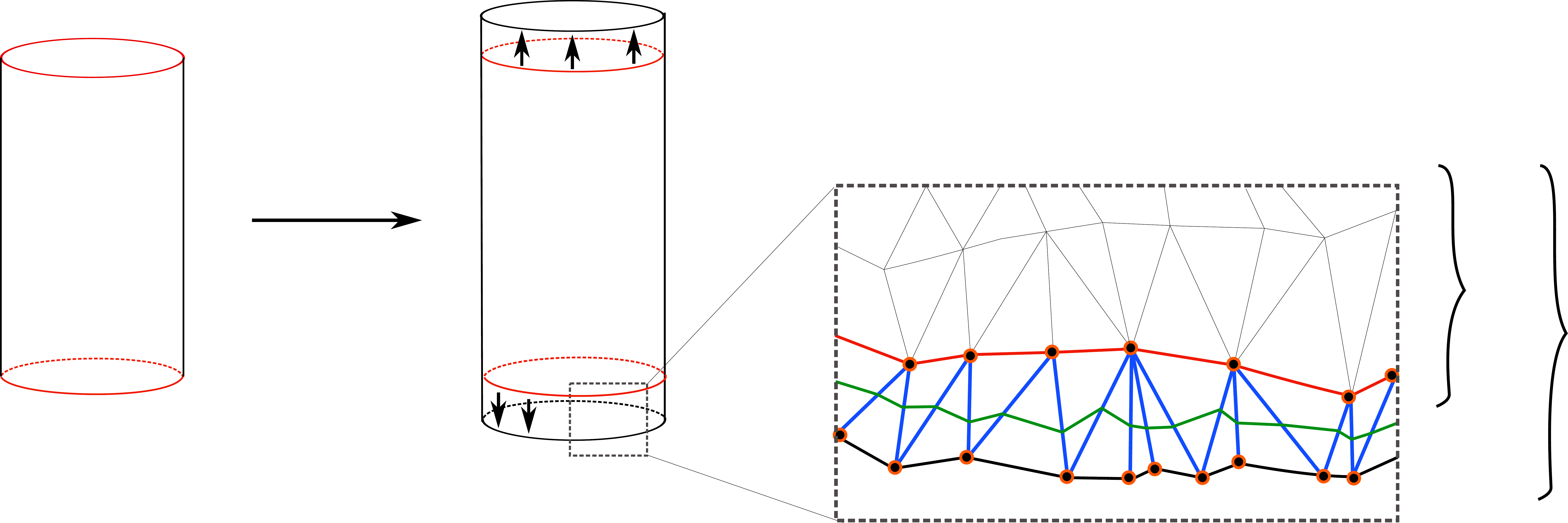}
    \put(-23,51){$M_i$}
    \put(3,40){$M^{\prime}_i$}
    \put(-370,-15){(a)}
    \put(-250,-15){(b)}
    \put(-90,-15){(c)}

    \caption{An illustration of the partition stage. (a) A connected component of the submanifold $M_i$ is obtained. (b) $M_i$ is extended to $M^{\prime}_i$ by adding the crossing edges. (c) A closeup of the crossing edges of the manifold $M^{\prime}_i$. The blue edges represent the crossings edges of $E_{i-1}$, and the green curve represents the portion of the curve $f^{-1}(c_{i-1})$ within the closeup region.}
    \label{fig:stage1}
\end{figure}

In this first stage of the parallel algorithm, we sort the vertices of the manifold with respect to the function $f$. This step can be done efficiently in parallel~\cites{tsigas2003simple,singler2007mcstl}. The critical sets for saddle points are then determined by assigning a thread to each saddle point. This computation is only necessary to determine the number of ascending paths that we need to initiate from that saddle point. Next, we partition the manifold $M$ to submanifolds along with certain regular values of the scalar function $f$.  More precisely, the partition stage is given as follows:

\begin{enumerate}

    \item Compute the critical points of $f$ by assigning a thread to each vertex in $M$. Let $p_1, p_2, ..., p_n$ be the list of critical points of $f$, and let $t_1, t_2, ..., t_n$ be their corresponding critical values. 

    \item We choose $k$ regular values $C=\{c_1,\cdots,c_k\}$ of $f$. These values will be utilized to slice the manifold $M$ into $k+1$ submanifolds $M_1$,...,$M_{k+1}$, such that the vertex counts of the submanifolds are approximately equal to each other. We also need to determine the connected components $f^{-1}(c_i)$ for $c_i$ in $C$. The connected components of the regular value $c_i \in C$ can be computed in linear time with respect to the number of edges in $M$ as follows. We visit all edges of $M$ and detect if an edge crosses one of the values $c_i$. If such an edge is found at a level $c_i$, then we keep rotating around to find all other edges crossing the value $c_i$ within the same connected component of $f^{-1}(c_i)$. After visiting all edges, we also  have determined the connected components of $f^{-1}(c_i)$ for each $c_i \in C$. We will denote the set of all connected components of $f^{-1}(c_i)$ for $1 \leq i \leq k $ by $\mathcal{C}_f$. We also call an edge in $M$ that crosses $f^{-1}(c_i)$ a \textit{crossing edge}. See Figure~\ref{fig:stage1} for an illustration. 

    \item Divide the surface $M$ into $k+1$ partitions along the level sets $f^{-1}(c_i)$ for all $c_i \in C$. We obtain a list of submanifolds $M_{[c_{0},c_1]}$,...,$M_{[c_{k},c_{k+1}]}$. Here we set $c_0=t_1$ and $c_{k+1}=t_n$. We will denote to $M_{[c_{i-1},c_{i}]}$ by $M_{i}$. The set $\{ M_i |1 \leq i \leq k+1 \}$ will be denoted by $\mathcal{M}_f$. 

    \item Next, we extend $M_i$ by adding other vertices in $M$ as follows. Let $E_{i-1}$ and $E_{i}$ be the edges in $M$ that intersect with $f^{-1}(c_{i-1})$ and $f^{-1}(c_{i})$, respectively. The submanifold $M^{\prime}_i=M^{\prime}_{[c_{i-1},c_{i}]}$ is obtained from $M_{[c_{i-1},c_{i}]}$ by adding the vertices from $E_{i-1}$ and $E_{i}$. We call the edges $E_{i-1}$ and $E_{i}$ the boundary edges of $M^{\prime}_{[c_{i-1},c_{i}]}$. Note that every two consecutive submanifolds from $ \mathcal {M}_{f}$ intersect with each other along their boundary edges. See Figure~\ref{fig:stage3} for an illustration. 
\end{enumerate}

The purpose of extending the submanifolds $M_{i}$ to $M^{\prime}_{i}$ in step (3) will be justified in the gluing stage in Section~\ref{the gluing stage}.

\subsection{Computing the Reeb Graph of Each Submanifold}
\label{second stage}
The manifold $M^{\prime}_i$ is, in general, homeomorphic to the manifold $M_i$, since the former is obtained from the latter by extending its boundary slightly. However, in the piece-wise linear case, the extension specified in the previous section may change the topology of the manifold. This will occur when  a crossing edge in $E_i$ or $E_{i-1}$ contains a critical vertex of $f$. When this case occurs, we exclude this vertex from $M^{\prime}_i$ in order to keep it homeomorphic to $M_i$. Using this convention, we can assume that the Reeb graph of the restriction of $f$ on $M_{i}$ is identical to the Reeb graph of the restriction of the Reeb graph of $f$ on $M^{\prime}_{i}$.

Computing the Reeb graphs $R(M^{\prime}_i,f)$ for $M^{\prime}_{i}$, for $1\leq i \leq k+1 $, now goes as follows. To each connected component in $M^{\prime}_{i}$ for $1\leq i \leq k+1 $, we assign a thread and the Reeb graph $R(M^{\prime}_i,f)$ on the submanifold $M^{\prime}_{i}$, which can be computed concurrently.

An ascending path that starts at a crossing edge or ends at a crossing edge will be treated specially. We call the Reeb graph edge that corresponds to such an ascending path a \textit{crossing arc}. Furthermore, if the starting or the ending edge of this ascending path is a crossing edge, then we will call the corresponding node in the Reeb graph a \textit{crossing node}. Every crossing node is determined by its crossing edge. In other words, given a crossing edge, we can recover its crossing node in the graphs $R(M_i^{\prime},f)$ for $1 \leq i \leq k+1$. In practice, we need to be able to do this in constant time, so we create a global map $G$ that takes as an input a crossing edge and returns its corresponding crossing node. In the case when a single crossing edge is associated with two crossing nodes from two consecutive submanifolds, then the map $G$ associates that edge crossing edge to the two crossing nodes (this occur when the ending edge of a crossing arc and the starting edge of the crossing arc in the consequent submanifold are the same). If the crossing edge is not associated with any crossing node, then this map returns a constant value indicating that this edge is not a starting or an ending crossing edge for any ascending path.  Note that each connected component in $\mathcal{C}_f$ has either two crossing edges that are associated with two crossing nodes or a single crossing edge that is associated with two crossing nodes. This map will be utilized in the gluing stage.

\subsection{The Gluing Stage}
\label{the gluing stage}

\begin{figure}[!ht]
    \centering
    \includegraphics[width=0.8\textwidth]{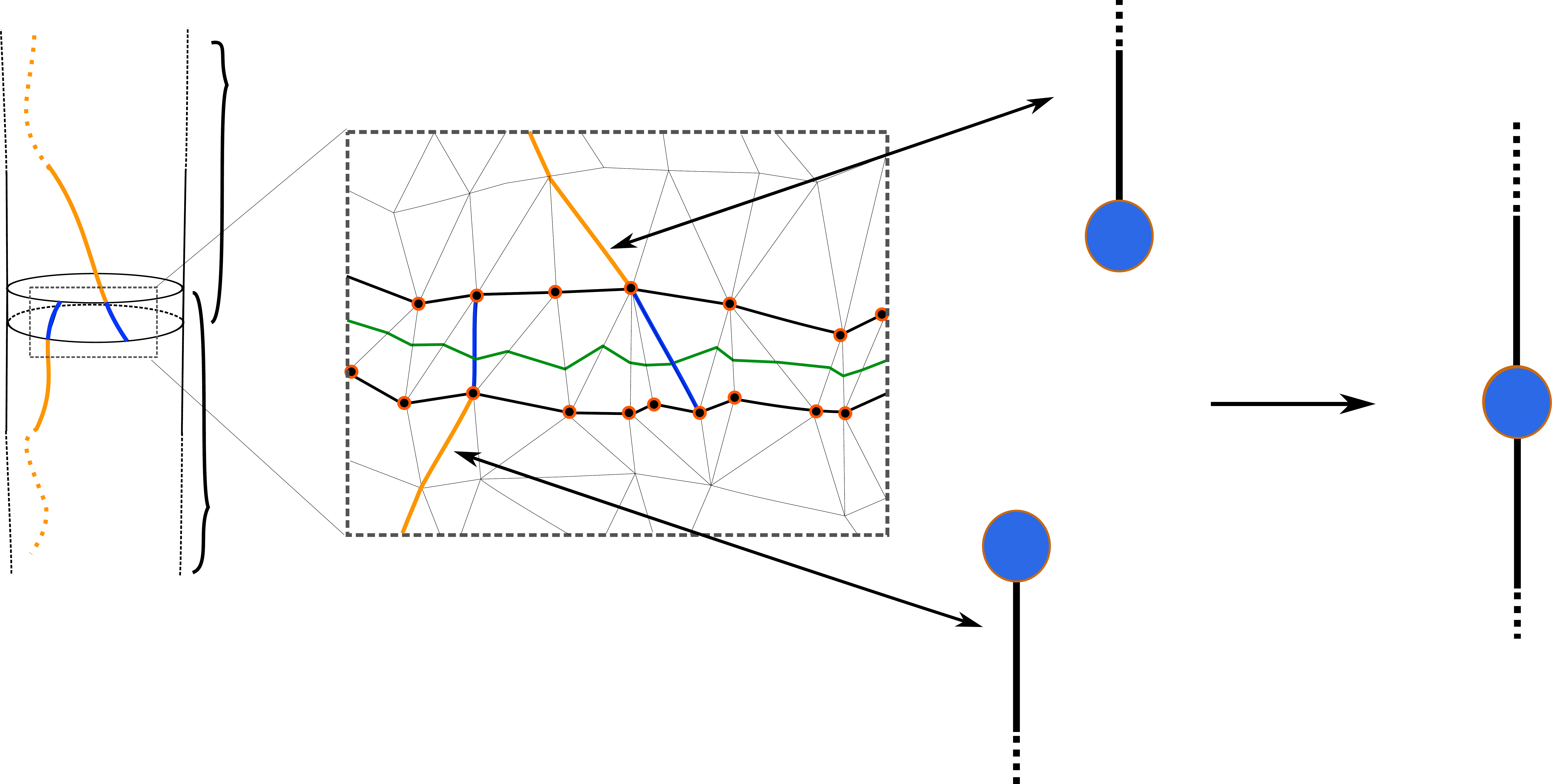}
    \put(-90,140){\small{$R(M_2,f)$}}
    \put(-120,40){\small{$R(M_1,f)$}}
    \put(-319,168){\small{$M_2$}}
    \put(-325,60){\small{$M_1$}}
    \put(-370,-15){(a)}
    \put(-250,-15){(b)}
    \put(-10,-15){(c)}

    \caption{An illustration of the gluing stage. (a) At this stage, two ascending paths from all submanifolds have been calculated. In the case when two consecutive submanifolds share a crossing edge, such as the case in the illustrative figure, the ascending path of the lower submanifold $M_1$ gets terminated at the at a crossing edge. Similarly, an ascending path from the higher submanifold $M_2$ gets initiated at a crossing edge. (b) A zoomed version of Figure~(a) shows the two crossing edges that determine uniquely two nodes in the Reeb graphs $R(M_1,f)$ and $R(M_2,f)$. The fact that these two edges belong to the same connected component in the inverse image of the regular crossing value is used to glue the graphs $R(M_1,f)$ and $R(M_2,f)$ along the blue nodes to obtain the graph in Figure~(c).}
    \label{fig:stage3}
\end{figure}
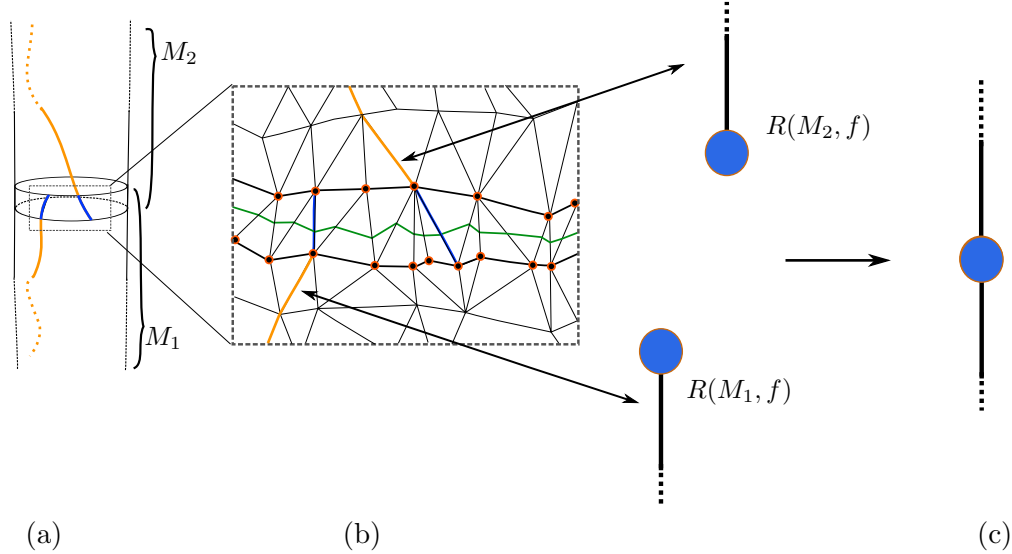

For the gluing stage, we need to glue the nodes of the Reeb graphs that occur in duplication $C_f$. For this purpose, we utilize the function $G$ that we constructed in Section~\ref{second stage}. For each connected component in $\mathcal{C}_f$, we visit its crossing edges and check if two edges within that connected component have been flagged by $G$. If this is the case, then we retrieve the crossing nodes that correspond to these two edges via the function $G$ and glue them. In the case when a connected component of $\mathcal{C}_f$ has a single crossing edge, then we retrieve the two crossing nodes that are associated with that edge and glue them. When we finish visiting all connected components of $\mathcal{C}_f$, all duplicate nodes will have been glued, and the final graph is $R(M,f)$. See Figure~\ref{fig:stage3} for an illustration of this process.

\section{Augmented Reeb Graph Computations: Going From the Reeb Graph to the Manifold}
\label{Reeb graph to Manifold}

In this section, we give an algorithm that describes an explicit computation of the map  $F: R(M,f) \longrightarrow M$. The computation of this map, alongside the computation of the Reeb graph, is usually called the augmented Reeb graph~\cite{harvey2010randomized}. Our algorithm here is the first parallel augmented Reeb graph algorithm that we are aware of. This map associates to every vertex $v$ in the graph $R(M,f)$ to the critical set $C_v$ associated to that critical point. More importantly, for each interior point $p$ of an edge in $R(M,f)$, we want to associate a simple closed curve $Cr_p$ in $M$ such that $\pi(Cr_p)=p$.

\subsection{Building the augmentation map $F: R(M,f) \longrightarrow M$ }
\label{map}
In the construction above, for our Reeb graph algorithm, an edge of the graph $R(M,f)$ is traced as a sequence of edges on the mesh running between two critical sets of the function $f$. This immediately gives an embedding of the edges of the Reeb graph on the surface. This embedding is used to get the circle corresponding to any points on the graph. More precisely, we have the following correspondence. Let $e$ be an edge of the graph $R(M,f)$. By the construction of our algorithm, every edge in $R(M,f)$ is determined by two critical points and a sequence of oriented edges on the mesh. Let $\mathcal{E}_e:=\{E_1,\cdots,E_n\}$ be the sequence of oriented edges on the mesh $M$ that corresponds to the edge $e$. If $E_i=[v_i,v_{i+1}]$, then we will denote by $l_f(E_i)$ to $|f(v_i)-f(v_{i+1})|$. Let $T_j(\mathcal{E}_e)$ be the summation $\sum_{i=1}^j l_f(E_i)$, where $1\leq j\leq n$.

\begin{figure}[!ht]
    \centering
    \includegraphics[width=0.9\textwidth]{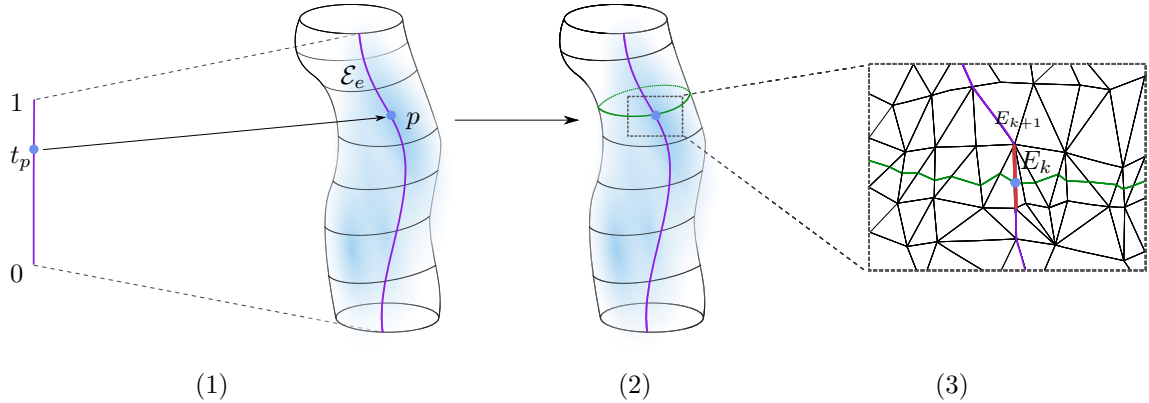}
    \put(-430,35){\small{$0$}}
    \put(-430,100){\small{$1$}}
    \put(-430,80){\small{$t_p$}}
    \put(-280,95){\small{$p$}}
    \put(-305,110){\small{$\mathcal{E}_e$}}
    \put(-48,78){\small{$E_k$}}
    \put(-58,95){\tiny{$E_{k+1}$}}
    \put(-360,-7){\small{$(1)$}}
    \put(-200,-7){\small{$(2)$}}
    \put(-80,-7){\small{$(3)$}}
    \caption{An illustration of the data retrieval procedure. (1) A value $t_p$ is selected from the interval $(0,1)$. (2) This value is used to determine a point $p$ on the Reeb graph edge $\mathcal{E}_e$. (3) This point is contained in an edge $E_k$ on the mesh $M$. This edge is used to determine the circle $Cr_p$ }
    \label{reebgraph2 mesh}
\end{figure}

An interior point $p$, on the edge $e$, is specified by giving a value $t_p$ in the interval $(0,1)$. To obtain the circle $Cr_p$ on the mesh $M$ that correspond to $p$, we do the following procedure:

\begin{enumerate}
    \item Map the interval $(0,1)$ linearly to the interval $(0,T_n(\mathcal{E}_e))$.
    
    \item Use the constructed linear function computed in step (1)  to map $t_p$ to its corresponding value $t^{\prime}$ in $(0,T_n(\mathcal{E}_e))$.
    
    \item Determine the edge $E_k=(v_k,v_{k+1})$ in $\mathcal{E}_e$ such that $T_{k-1}(\mathcal{E}_e)<t^{\prime}\leq T_{k}(\mathcal{E}_e)$.

    \item Now we need to find the value $t$, in the range of $[a,b]$, the range of $f$, such that $f^{-1}(t)$ contains $Cr_p$. We know that this value corresponds to $t^{\prime}$, which lies in the interval $[T_{k-1}(\mathcal{E}_e),T_k(\mathcal{E}_e)]$. The required $t$ lies in interval $[f(v_k),f(v_{k+1})]$. Hence, we map the interval $[T_{k-1}(\mathcal{E}_e),T_k(\mathcal{E}_e)]$ linearly to $[f(v_k),f(v_{k+1})]$ and determine the value $t$ in $[f(v_k),f(v_{k+1})]$ that corresponds to $t^{\prime}$.     
    
    \item The required circle $Cr_p$ is precisely the connected component of $f^{-1}(t)$ that contains the edge $E_k$.
\end{enumerate}
See Figure~\ref{reebgraph2 mesh} for an illustration of the main parts of the previous procedure.

Note that in the previous procedure we rely on the fundamental assumption that $f$ takes different values on the vertices of the mesh. 

\subsection{Consistent Parameterization of The Edges of the Reeb Graph}

\begin{figure}[h]
\tiny{
  \centering
   {\includegraphics[scale=0.15]{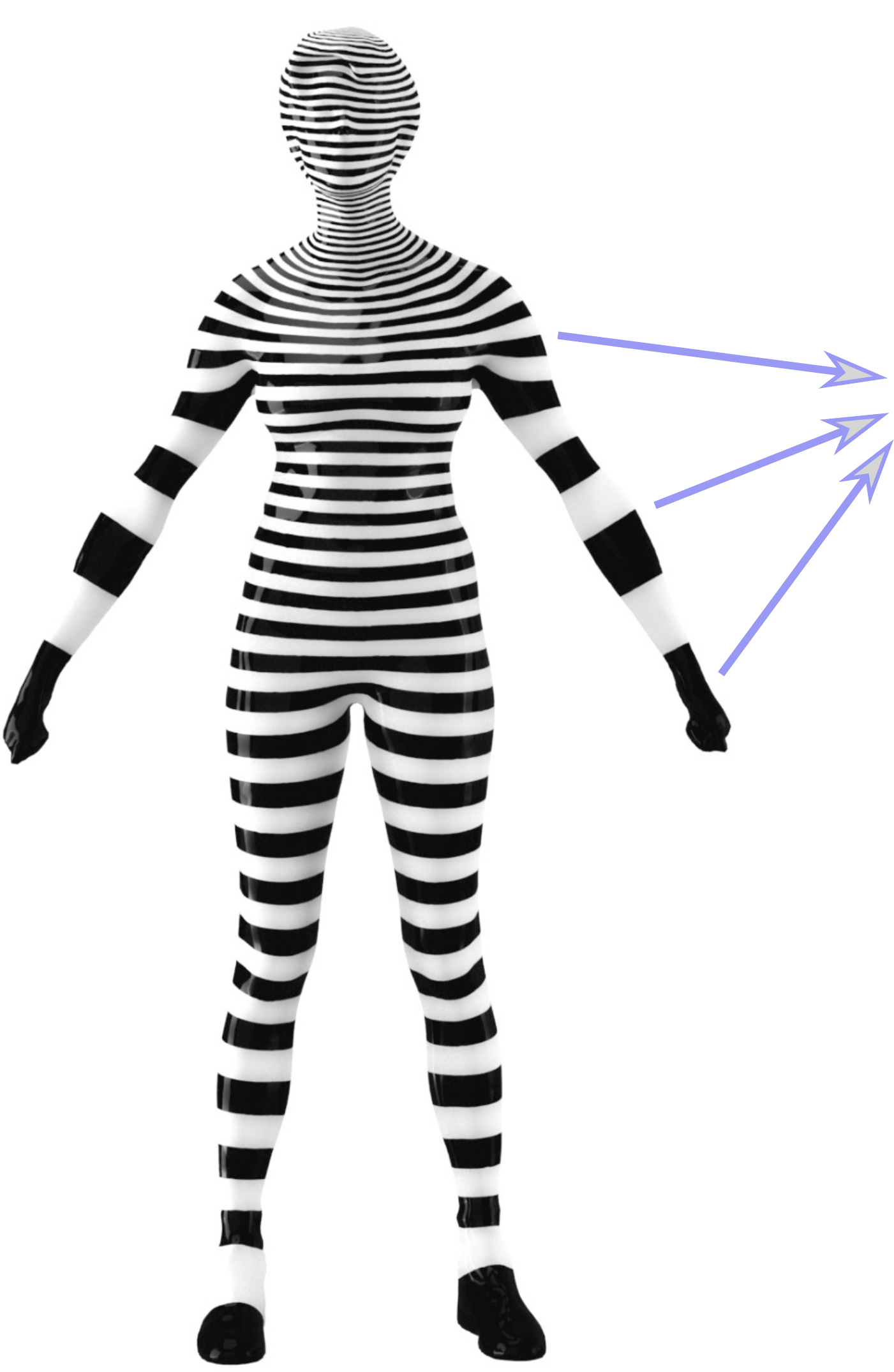}
 \put(5,105){different gradient values of $f$}}
     \vspace{.2in}
     \caption{The arm of the female character mesh corresponds to a single arc in the Reeb graph associated with the scalar function indicated on the mesh. Observe that the gradient of this scalar function varies along this arm which makes it hard to parametrize the mapping between the Reeb graph and the corresponding regions on the mesh.}
     \label{problem}}
\end{figure}

Going back to \ref{map} we observe that the mapping between $(0,1)$ and $(0,T_n(\mathcal{E}_e))$ depends on the gradient of the function $f$. More specifically, recall that $T_j(\mathcal{E}_e) = \sum_{i=1}^j l_f(E_i)= \sum_{i=1}^j |f(v_i)-f(v_{i+1}) |$. Hence the values of $l_f$ for a given edge $E$ can be interpreted as the gradient of $f$ along that edge. Hence the derivative of the function $j \to T_j(\mathcal{E}_e)$ in not constant in general. This variability in the derivative can make it difficult and less intuitive to choose a value $t$ in $(0,1)$ that corresponds to a specific curve on the mesh because, while the mapping between $(0,1)$ is linear, the function $j \to T_j(\mathcal{E}_e)$ is not linear in general. See Figure \ref{problem} for an illustration of the gradient problem.

 To this end for a given function $f$ we want to construct another function $\hat{f}$ that has the following two properties :
 \begin{enumerate}
     \item The levels sets of the function $\hat{f}$  are parallel the level sets of $f$.
     \item The function $\hat{f}$ has uniform gradient everywhere. 
 \end{enumerate}
 We need to recall the definitions of gradient and divergence of a triangulated mesh quickly before we give the construction of $\hat{f}$. 

\subsubsection{Gradient and Divergence of a PL function on a triangulated Mesh}
\begin{figure}[h]
  \centering
   {\includegraphics[scale=0.4]{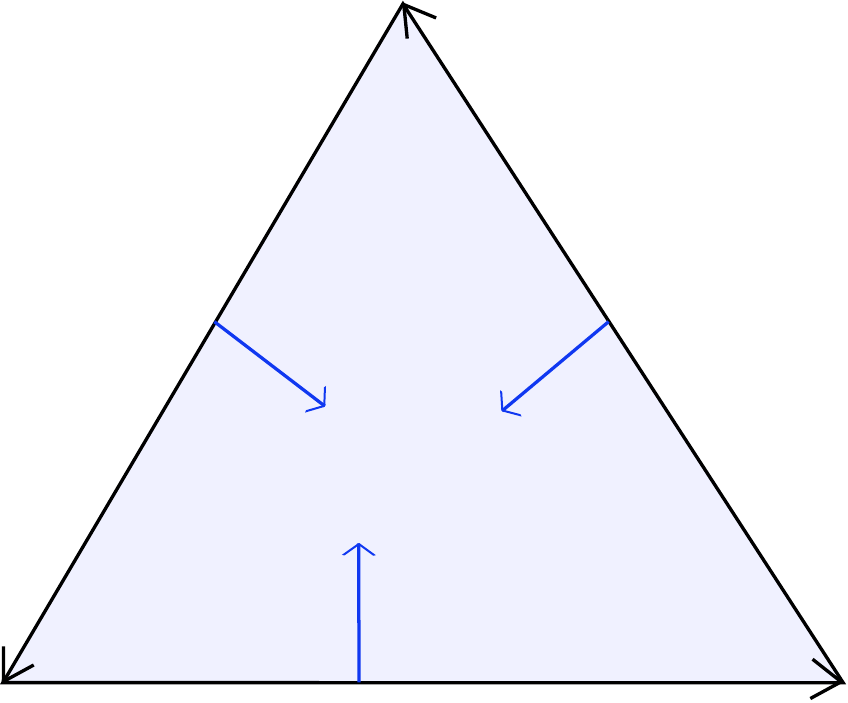}
   \put(-90,138){$v_i$}
   \put(-5,-7){$v_{i-1}$}
   \put(-18,7){$\beta_{i}$}
   \put(-170,-7){$v_{i+1}$}
   \put(-154,6){$\alpha_{i}$}
   \put(-40,-8){$E_i$}
   \put(-30,50){$E_{i+1}$}
   \put(-160,50){$E_{i-1}$}
   \small{
   \put(-90,10){$\nabla B_{i}$}
    \put(-70,46){$\nabla B_{i+1}$}
    \put(-120,46){$\nabla B_{i-1}$}
    }
  	\caption{The gradients of hat functions of a triangle.}
    \label{bary}
 }
\end{figure}

 Let $f$ be a PL scalar function on a triangulated manifold $M$. Let  $Face=[v_i,v_{i+1},v_{i+2}]$ be a face in $M$. Denote by $E_i$ be the counterclockwise oriented edge opposite to the vertex $v_i$. See Figure \ref{bary}. Let $B_i : Face\to \mathbb{R}$ be the hat function on the vertex $v_i$ defined by $B_i(v_j ) = \delta_{ij} $ for
$i, j = 1, 2, 3$. The gradient of $f$ is a constant and  tangential vector on $Face$ given by ~\cite{pinkall1993computing} : 

\begin{equation*}
\nabla f (Face) = \sum_{i=1}^3 \nabla B_i f(v_i),    
\end{equation*}
where 
\begin{equation*}
\nabla B_i=\frac{||E_i||} {2A_{Face}}\overrightarrow{u_i} 
\end{equation*}
Here $\overrightarrow{u_i}$ is a unit vector perpendicular to the vector $E_i$ and oriented so that it points into the face $Face$ and $A_{Face}$ is the area of the face $Face$. See Figure \ref{bary}.

\begin{figure}[h]
  \centering
   {\includegraphics[scale=0.22]{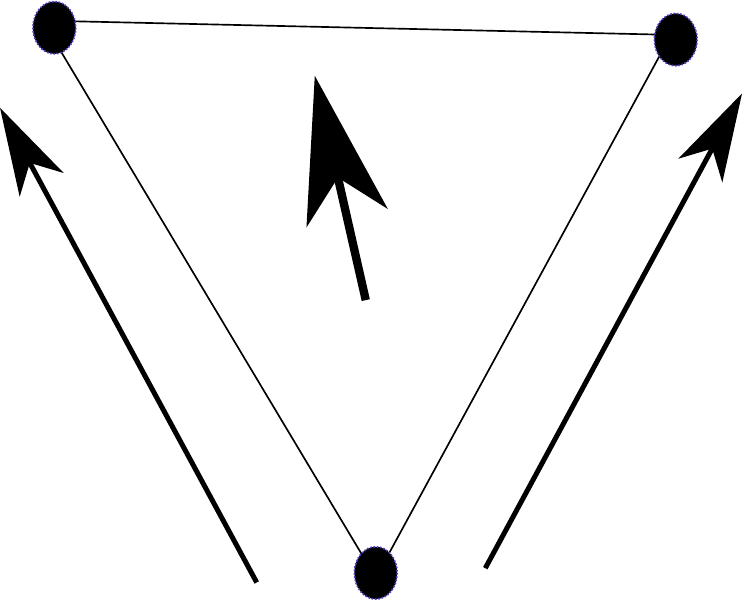}
   \small{
   \put(-40,-7){$v_i$}
   \put(-26,51){$\theta_{j_1}$}
    \put(-65,50){$\theta_{j_2}$}
    \put(-39,33){$X_j$}
    \put(-72,15){$e_{j_1}$}
     \put(-16,15){$e_{j_2}$}}
  	\caption{Computing the divergence at vertex $v_i$}
    \label{il}
 }
\end{figure}

The divergence of a vector field $X$ defined on the vertices on $M$ was given in ~\cite{polthier2002polyhedral} and it can be computed via the formula: 
\begin{equation*}
 {\rm div}\ X (v_i)=\frac{1}{2}\sum_{j \in  F(i)} \cot \theta_{j_1} \left\langle  e_{j_1},X_j \right\rangle +\cot \theta_{j_2} \left\langle  e_{j_2},X_j \right\rangle,
\end{equation*}
where $F(i)$ is the set of indices of all faces that are incident to the vertex $v_i$,  $e_{j_1}, e_{j_2}$ are the two vectors in face $j$ that contain the vertex $v_i$ and $\theta_{j_1}, \theta_{j_2}$ are the angles that are opposite the edges $e_{j_1}$ and $e_{j_2}$ respectively. See Figure \ref{il}.

\subsubsection{Unit Gradient Scalar Fields}
Now given a function  $f:M \longrightarrow \mathbb{R}$ on a triangulated mesh $M$, we are interested in finding a function $\hat{f}$ with the following two properties: 
 \begin{enumerate}
     \item The levels sets of the function $\hat{f}$  are parallel to the level sets of $f$.
     \item The function $\hat{f}$ has uniform gradient everywhere. 
 \end{enumerate}
The above two conditions are equivalent to constructing a function $\hat{f}$ such that ${\rm grad}\ \hat{f} \approx \frac{{\rm grad}\ f}{||{\rm grad}\ f||}  $. Setting $X:=\frac{{\rm grad}\ f}{||{\rm grad}\ f||} $ finding such $\hat{f}$ can be obtained by solving the following Poisson equation :
\begin{equation*}
\min_{\hat{f}} \int_M  | \nabla \hat{f} -X|dM,
\end{equation*}      
which is equivalent to solving $\Delta \hat{f}={\rm div}\ X$ where $\Delta$ is the Laplacian of the mesh $M$. Hence finding the function $\hat{f}$ can be reduced to the following two simple steps :

\begin{enumerate}
\item Compute $X = \frac{{\rm grad}\ f}{||{\rm grad}\ f||}.$    
\item Solve the Poisson equation $\Delta \hat{f}={\rm div}\ X$.
\end{enumerate}
The previous algorithm works on any generic function $f$ such that the gradient of $f$ is not zero. Moreover, Step (2) can be easily solved using the definitions of the gradient and the divergence provided earlier. The above simple procedure is a generalization of the geodesic in heat method \cite{crane2013geodesics} where the desired function $\hat{f}$ represents a distance function.

\begin{figure}[h]\centering
  \includegraphics[scale=.13]{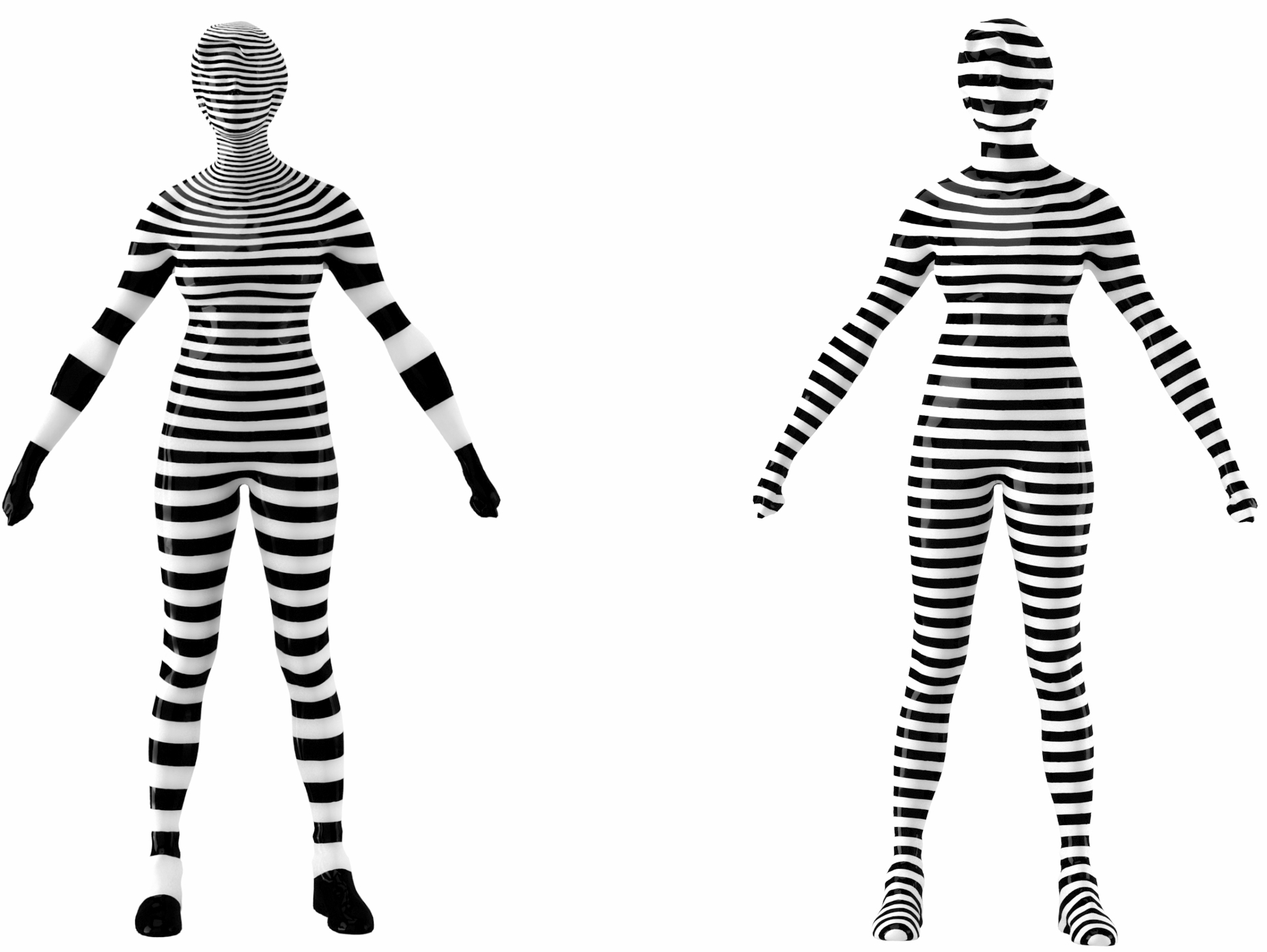}
  \caption{Obtaining a unit gradient scalar field. }
  \label{poisson compare}
\end{figure}

An example of such a procedure is illustrated in Figure \ref{poisson compare}. The right model in the figure shows a solution for a scalar function $\hat{f}$ obtained as a Poisson equation $\Delta  \hat{f}= {\rm div}\ X $  where $X$ is the normalized gradient of the original function $f$ shown on the left. Observe that the level sets of both $f$ and $\hat{f}$ are parallel but now the gradient of $\hat{f}$ is constant.

\section{Run-time Analysis and Implementation}

We tested the presented algorithm on meshes with various complexities. In particular, we performed the speedup analysis of the parallel algorithm to our implementation of the sequential version given \cite{doraiswamy2008efficient}.   Our experiments were done on an AMD FX 6300 6-Core with 32 GB memory. The algorithm was  implemented in C++, and the Windows platform was used.

\begin{figure}[!ht]
    \centering
    \includegraphics[width=0.8\textwidth]{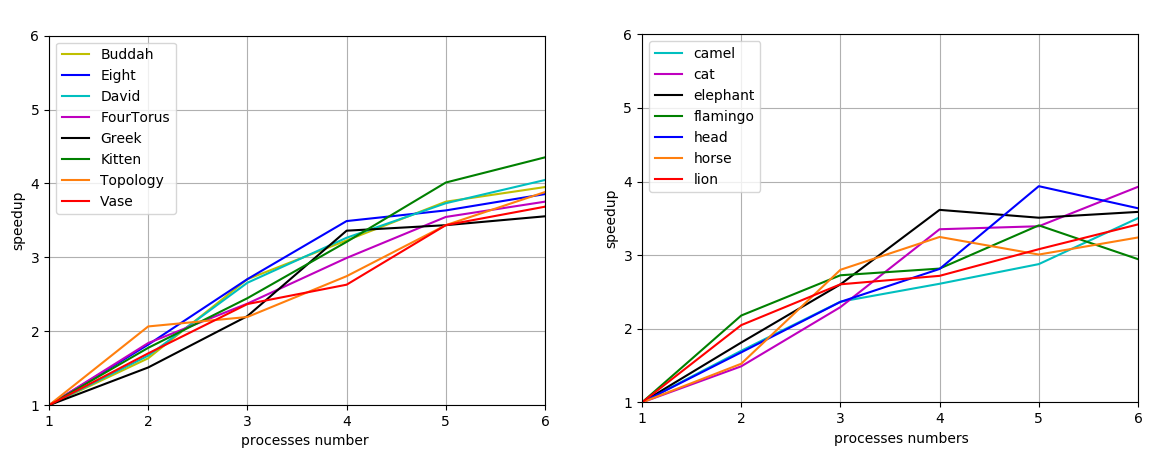}
    \caption{Speedups obtained by our parallel Reeb graph algorithm.} 
    \label{perf}
\end{figure}

We test our parallel algorithm with two datasets: the AIM@SHAPE Repository as well as the MeshDeform dataset available in \cite{sumner2004deformation}. The initial attempt did not give us an increase of performance over the sequential algorithm for most meshes available in the above datasets.    
In order to take advantage of the parallel implementation, we tested our algorithm on high resolution meshes. Specifically, we uniformly increase the resolution of the meshes available these two datasets to $200k$. On the AIM@SHAPE library our implementation gave us a minimum speedup of 3.6, a maximum one of 4.3, and an average speedup of 3.8 on 6 cores. Using this, we obtain a $63\% $ average parallel efficiency. One the MeshDeform dataset we obtained a minimum speedup of 2.9, a maximum one of 3.9, and an average speedup of 3.5 on 6 cores. This dataset gives a $42 \%$ average parallel efficiency on 6 cores. The details are given in Figure~\ref{perf}. The $x$-axis represents the number of processes, and the $y$-axis shows the speedup. 

\begin{figure}[!ht]
    \centering
    \includegraphics[width=0.9\textwidth]{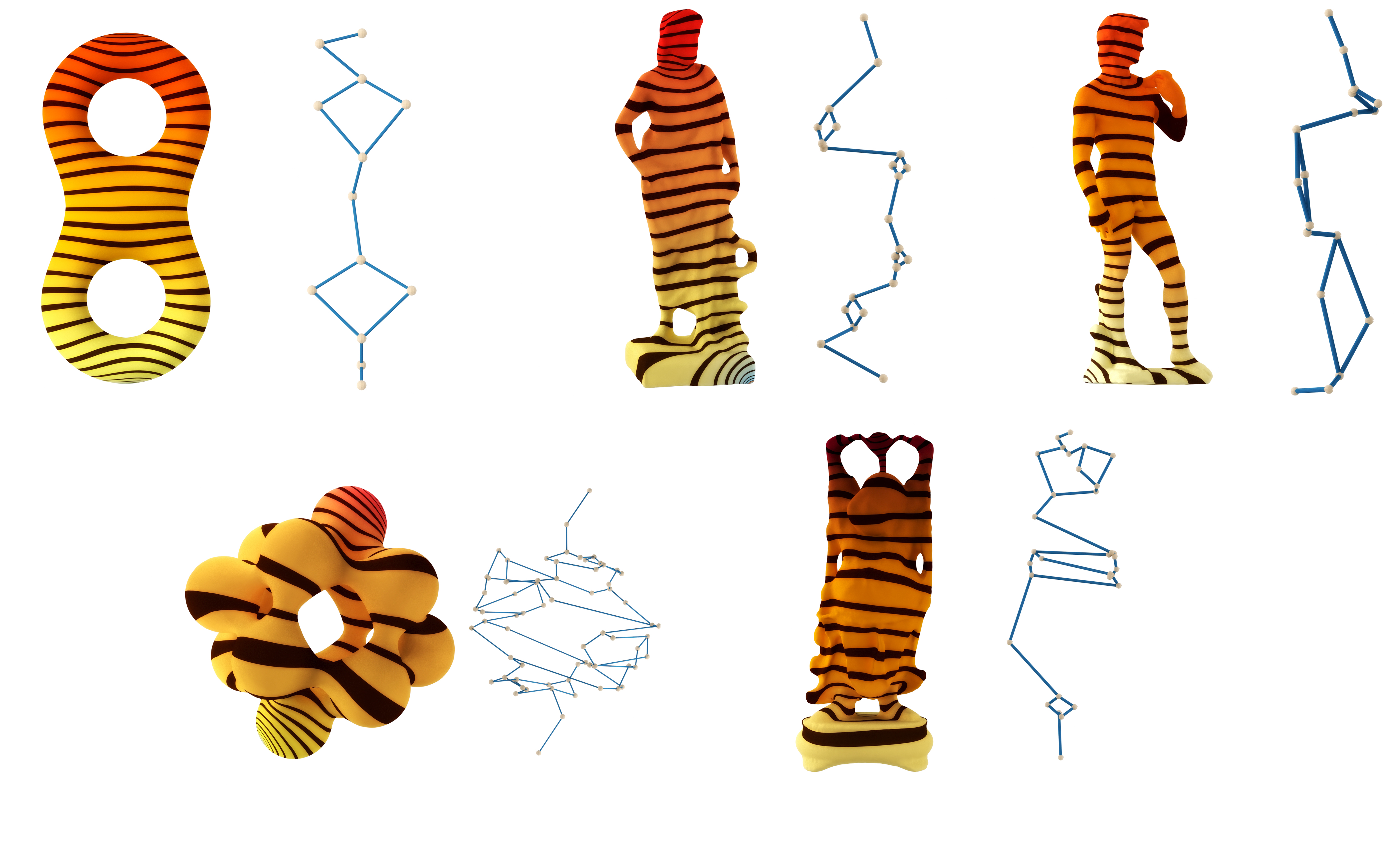}
    \caption{Some of the meshes that we used for our parallel Reeb graph computations.} 
    \label{allreeb}
\end{figure}

Figure~\ref{allreeb} shows a few examples of the meshes we utilized in our tests above, along with their corresponding Reeb graphs.

\section{Applications}
\label{application}

Reeb graphs on a surface $M$ encodes a rich amount of topological information from the original surface. In this section, we show how the Reeb graph of a surface $M$ gives rise to a natural collection of simple closed curves on $M$. The applications that we present can be described in terms of these curves. We first define these curves and show their relation with the Reeb graph. We then give a procedure to extract them by utilizing the tools we described in Section~\ref{Reeb graph to Manifold}. Finally, the curves are  utilized to obtain two higher genus surface mesh segmentation algorithms. Other Reeb graph-based segmentation algorithms can be found in~\cites{xiao2003discrete,xiao2003topological,tierny2007topology}.

Let $R(M,f)$ be a surface $M$ and a scalar function $f$. The edges of the Reeb graph $R(M,f)$ determine the following types of simple closed curves on $M$:
\begin{itemize}
    \item Cutting system curves.
    \item Pants decomposition curves.
    \item Branch curves.
\end{itemize}
We describe how a Reeb graph on a surface can be used to realize these curves next.

\begin{figure}[!ht]
    \centering
    \includegraphics[width=0.7\textwidth]{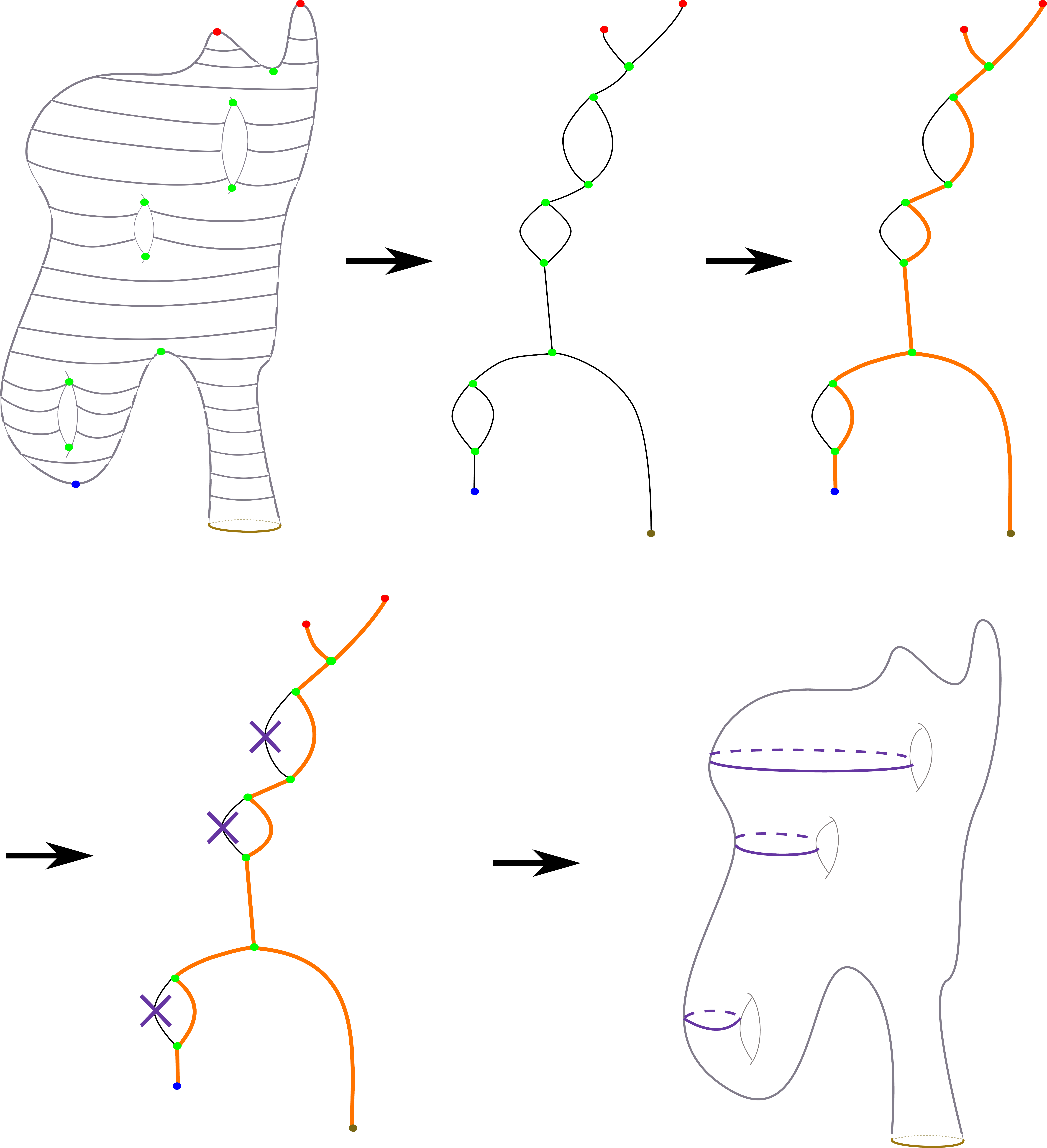}
    \caption{The steps of the cutting system algorithm. (1) mesh $M$ with a scalar function defined on it are given. (2) The Reeb graph $R(M,f)$ is computed. (3) We compute a spanning tree of $R(M,f)$. (4) We select all edges in the graph that do not belong the spanning tree we computed in (3) and then we select an interior point on each one of these edges. (5) For each interior point we used our augmented Reeb graph algorithm to compute the circle that corresponds to it  on the mesh. } 
    \label{cut system figure}
\end{figure}

\para{Cutting System Curves} 
Reeb graphs can also be used to determine the so-called \textit{cutting system}. A cutting system for a connected, closed, orientable surface $M$ of genus $g$ is a collection of unordered disjoint simple closed loops embedded in $M$ whose complement $M\backslash ( l_1\sqcup \cdots \sqcup l_g)$ is a sphere with $2g$ boundary components~\cite{hatcher1980presentation}. Segmenting a surface along a cutting system curve yields a genus zero surface with multiple boundary components. Hence this can be used to aid in mesh parametrization. See, for instance,~\cite{zeng2009generalized} and the references therein. We describe here a Reeb graph-based algorithm to obtain a cutting system. The algorithm is illustrated in the Figure \ref{cut system figure} and it goes as follows:  
\begin{enumerate}
    \item Let $T$ be a spanning tree of $R(M,f)$ and consider the edges $e_1,...e_g$ in $R(f)\backslash T$. 

    \item Select an interior point in $e_i$, for $0\leq i \leq g$.

    \item Each interior point selected in the previous step determines a loop $l_i$, which can be obtained using the Reeb graph algorithm we described here. 

\end{enumerate}

The steps of the cutting system algorithm are described in Figure \ref{cut system figure}.

\begin{figure}[!ht]
    \centering
    \includegraphics[width=0.95\textwidth]{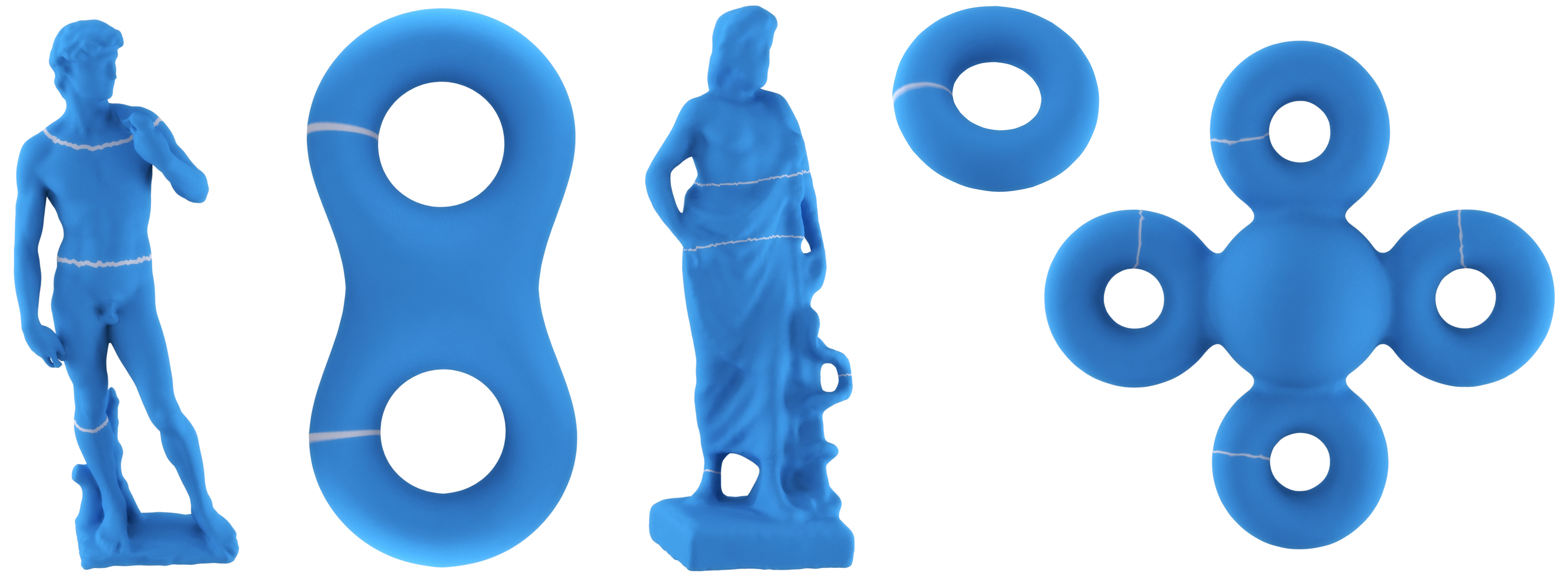}
    \caption{Examples of cut system curves on triangulated surfaces.} 
    \label{cut system}
\end{figure}

Figure~\ref{cut system} shows multiple examples of cutting system curves on triangulated surfaces.

\begin{figure}[!ht]
    \centering
    \includegraphics[width=0.7\textwidth]{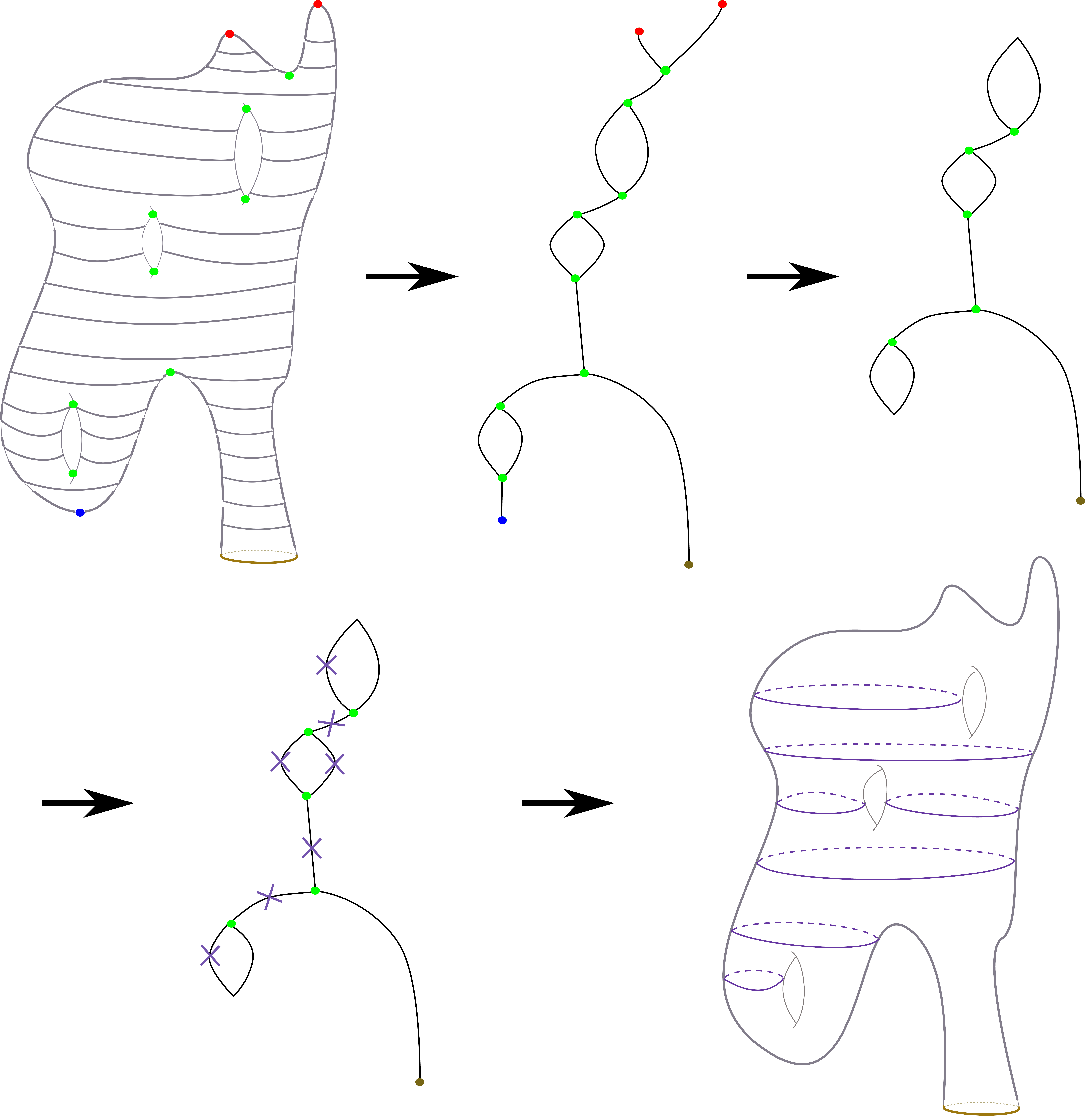}

    \caption{A Reeb graph can be used to segment a surface into a collection of topologically consistent patches. (1) The surface $M$, with a scalar function $f$, is given as an input. (2) The Reeb graph of $R(M,f)$ is calculated. (3) The deformation retract graph $R^{\prime}(M,f)$ of the graph $R(M,f)$ is calculated. (4) For each edge in the graph $R^{\prime}(M,f)$, we select a point. (5) We select the curves on the manifold $M$ that corresponds to the points selected in the previous step.} 
    \label{pants algorithm}
\end{figure}

\begin{figure}[!ht]
    \centering
    \includegraphics[width=0.75\textwidth]{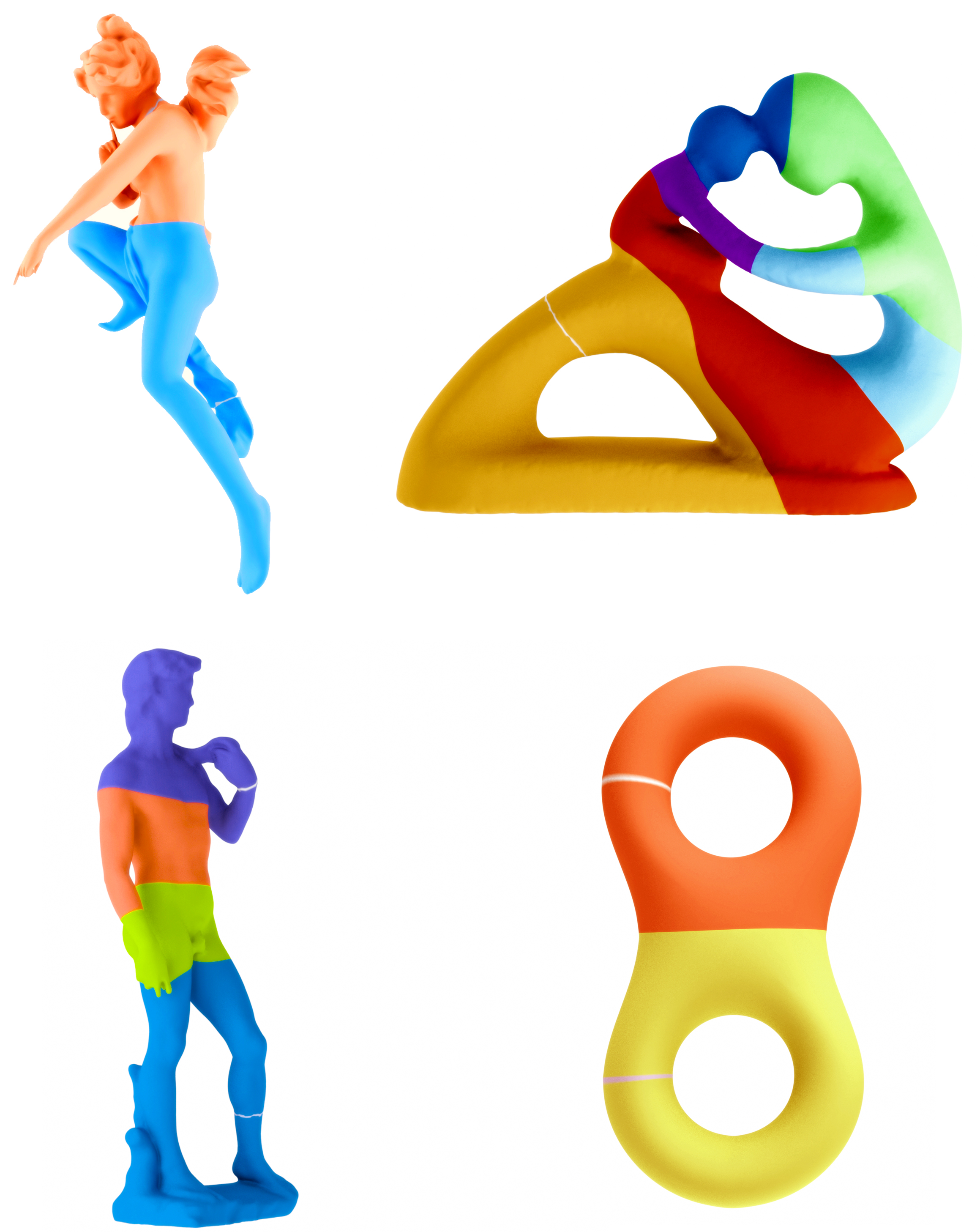}
    \caption{Examples of segmentation of higher genus surfaces into a pair of pants using our Reeb graph algorithm.} 
    \label{pants example}
\end{figure}

\para{Pants Decomposition Curves} 
A Reeb graph naturally gives rise to a collection of curves that can be used to decompose a surface into a pair of pants. A pair of pants is a genus zero surface with three boundary components. Beside surface segmentation~\cite{hajij2016segmenting}, surface pants decomposition has found applications in mesh parametrization~\cite{kwok2012constructing}, surface matching~\cite{li2009surface}, and surface classification and indexing~\cite{jin2009computing}. The method to obtain a pants decomposition from the Reeb graph is illustrated in Figure~\ref{pants algorithm} and is described as follows:

\begin{enumerate}
    \item Let $R^{\prime}(M,f)$ be the deformation retract of $R(M,f)$. This graph can be obtained by recursively deleting nodes with valency one from $R(M,f)$ and the edge attached to them until no such indices exist. We exclude from this deletion the $1$-valence nodes originating from the boundary of $M$. We also delete all nodes with valency $2$ and combine the two edges that meet at that node to form a single edge. This step is illustrated in step $(3)$ Figure~\ref{pants example}.

    \item We select one interior point from each edge in $R^{\prime}(M,f)$, provided this edge does not have a node of valency one. 
    
    \item We use our Reeb algorithm to determine the curves on the surface that correspond to the points that we selected on the graph in the previous step.
\end{enumerate}

The results of the previous algorithm were tested on triangulated meshes with various topological complexities. We show some examples in Figure~\ref{pants example}.

\para{Branch Curves} A branch curve on a surface $M$ is a simple closed curve that bounds a topological disk on $M$. Such a curve is also called null homotopic. A branch curve is determined by a Reeb graph edge, which has a vertex of valence one. Note that cutting the surface along a branch curve increases the number of connected components of the surface. Cutting along such curves can be used for the segmentation of a genus zero surfaces. There are many Reeb graph-based segmentation algorithms in the literature for segmentation of genus zero surfaces, such as the surface obtained from a humanoid character. See for instance~\cites{xiao2003discrete,xiao2003topological,tierny2007topology}. Since this type of curves is essentially utilized elsewhere in the literature, we simply list it here for completeness of our discussion. All these methods, however, lack the description of a method to extract the manifold data from the Reeb graph data.

\subsection{Choosing Morse Scalar Functions}
The cutting system and the pant decomposition curves that we presented in this section depend on the choice of the scalar function used to define the Reeb graph in the following way. In particular, the choice of these curves affects the final quality of the segmented mesh. For this reason it is important to choose a scalar function with certain desirable properties. By desirable properties we mean the following :

\begin{enumerate}
    \item The scalar function has a small number of critical vertices.
    \item The level sets of the scalar function follow the geometry of the mesh as closely as possible.
    \item The scalar function requires minimal input from the user.
\end{enumerate}
We briefly discuss several scalar functions with the above properties.

\subsubsection{Harmonic functions}

Condition $1$ is desirable because it leads in general to a simpler Reeb graph. For a surface mesh $M$ of genus $g$ one can always construct a scalar function $f$ on $M$ with $2g$ critical points. This can be done by the so called \textit{Harmonic functions.} Recall that a harmonic map on a triangulated surface $M$ is a scalar function $f:M \to \mathbb{R}$ that satisfies the Laplace equation $\Delta f=0$ subject to the Dirichlet boundary conditions $f(v_i)=c_i$ for all $v_i \in  V_C$. Here the set $V_C \subset V$ is a list of constrained vertices and $c_i$ have known scalar values providing the boundary conditions.  The reason for our interest in Harmonic functions is that they satisfy the so called \textit{maximum principle property} ~\cite{rosenberg1997laplacian} which asserts that the solution for the above system has no local extrema other than the constrained vertices. To achieve this setting in practice one has to be careful about the choice of the weights utilized to define the Laplacian. More specifically, for a triangulated mesh $M$ the standard discretization for the Laplacian operator at a vertex $v_i$ is given by :
\begin{equation*}
\Delta f(v_i)=\sum_{[v_i,v_j]\in M}w_{ij}(f(v_j)-f(v_i)), 
\end{equation*}
where $w_{ij}$ is a scalar weight assigned to the edge $[v_i,v_j]$ such that $\sum_{[v_i,v_j]\in M}w_{,j}=1$. Choosing the weights $w_{ij}$ such that $w_{ij} >0 $ for all edges $[v_i,v_j]$ guarantees the solution of the Laplace equation has no local extrema other than at constrained vertices $V_C$ ~\cite{floater2003mean}. These conditions are satisfied by the\textit{ mean value weights}:
\begin{equation*}
w_{ij}=\frac{tan(\theta_{ij}/2)+tan(\phi_{ij}/2) }{||v_j-v_i ||},
\end{equation*}
where the angles $\theta_{ij}$ and $\phi_{ij}$ are the angles on either sides of the edge $[v_i,v_j]$ at the vertex $v_i$. Mean value weights are used to approximate harmonic map and they have the advantage that they are always non-negative which prevents any introduction of extrema on non-constrained vertices in the solution of the Laplace equation specified above. Such a function can be obtained as a solution for Laplace equation with mean value weights and with only two constrained vertices $V_C=\{v_{min},v_{max}\}$ such that $f(v_{min})<f(v_{max})$. For instance all function shown in \ref{allreeb} are obtained by solving the Laplace equation with exactly two constrained vertices.

\subsubsection{The Poisson Equation}

\begin{figure}[h]
  \centering
   {\includegraphics[scale=0.22]{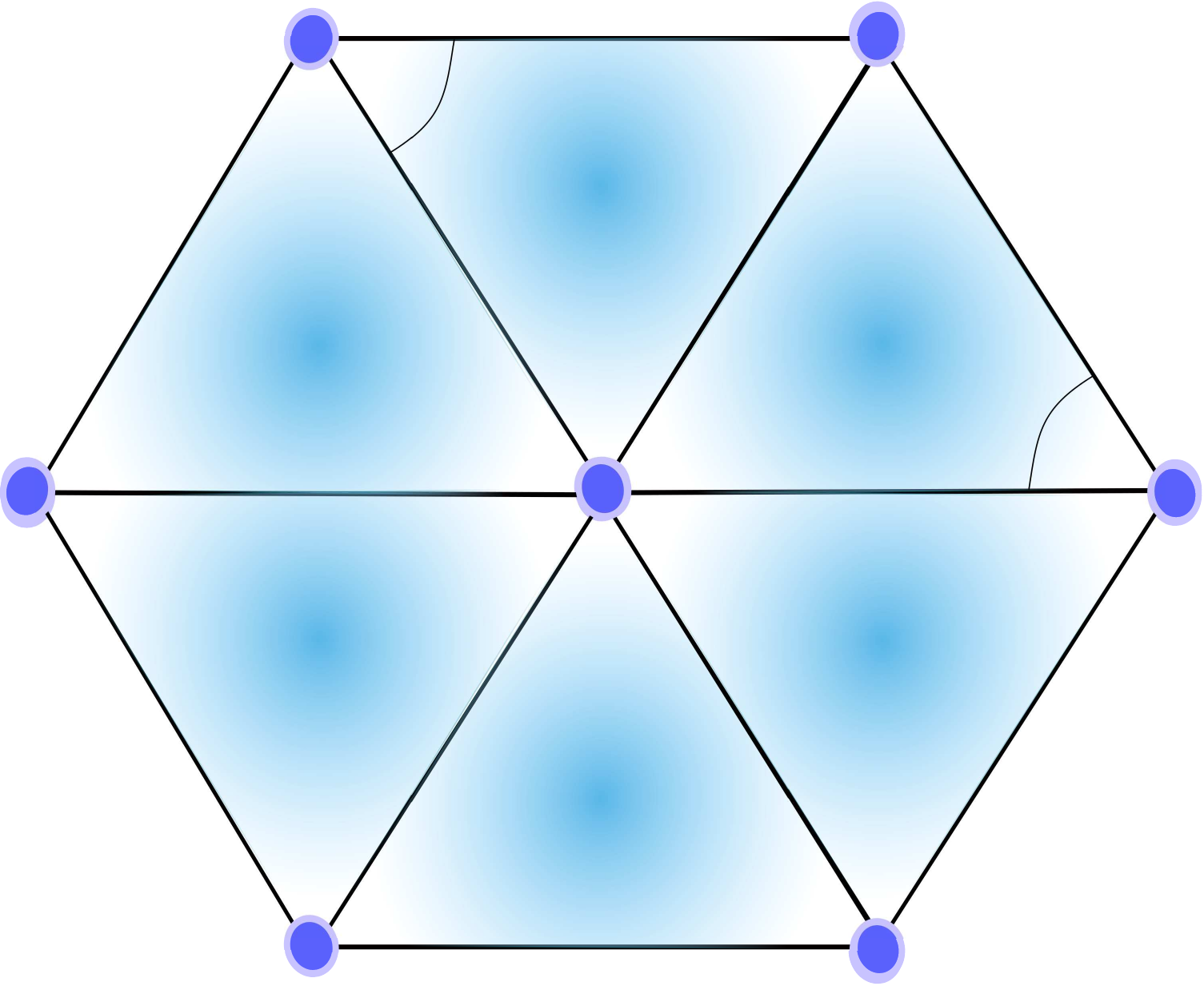}
   \small{
   \put(-84,74){$v_i$}
   \put(-26,71){$\theta_{ij}$}
    \put(-110,116){$\beta_{ij} $}
     \put(-32,118){$v_j$}}
  	\caption{The angles $\theta_{ij}$ and $\beta_{ij}$ are defined with respect to an edge $[v_i,v_j]$. }
    \label{ang}
 }
\end{figure}

\begin{figure}[!ht]
    \centering
    \includegraphics[width=0.4\textwidth]{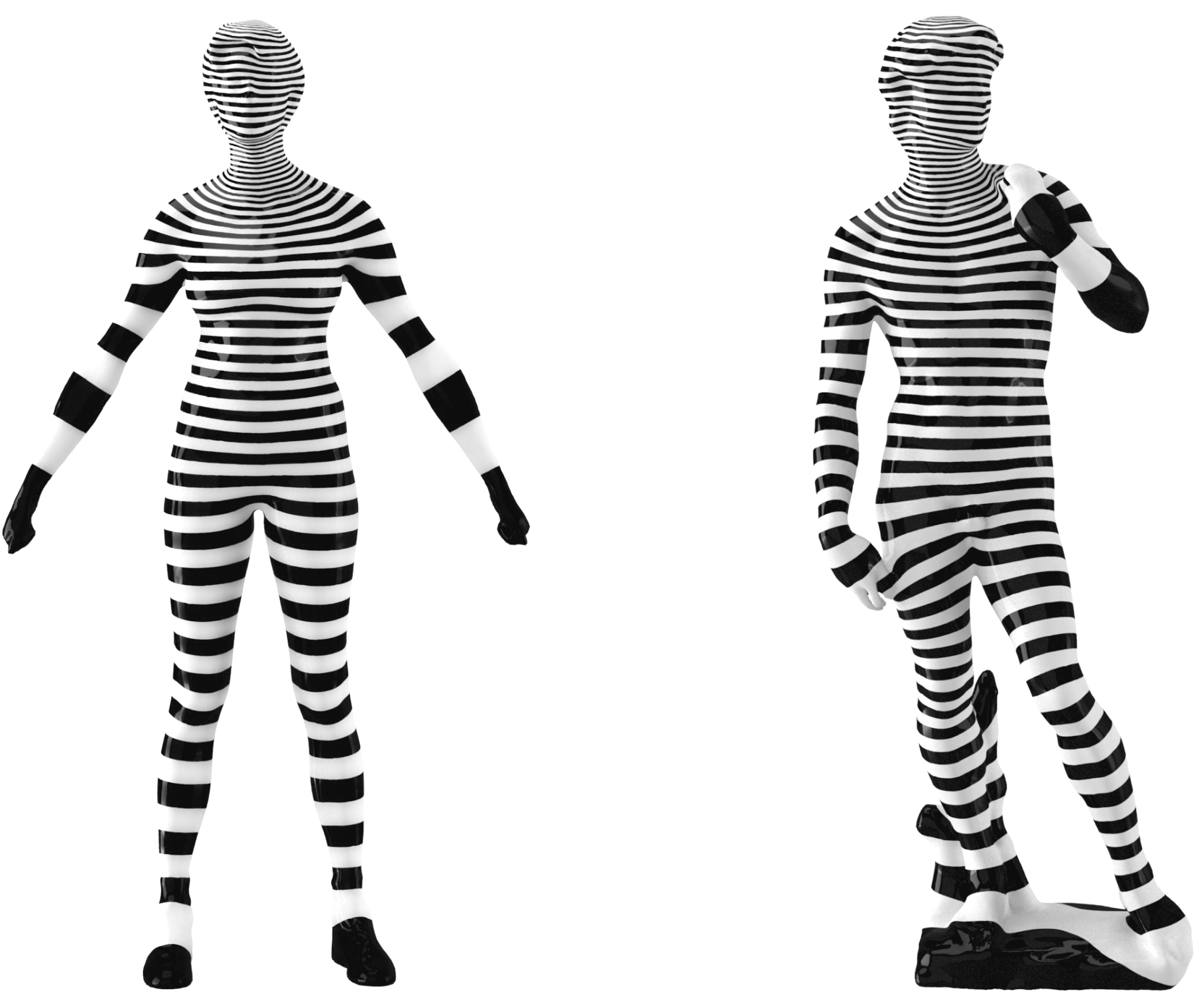}
    \caption{Example of Poisson fields on Triangulated meshes. In both characters, a single vertex, which is the highest node in the head of the character, was chosen to be the constrained vertex required to solve the Poisson equation. Observe how this scalar function follows the geometry of the mesh.} 
    \label{poisson123}
\end{figure}

The Poisson equation on a triangulated mesh with Dirichlet boundary condition is defined by:
\begin{equation}
\label{pois}
\Delta f = h, \quad f(v_i)=c_i  \textrm{ where }  v_i \in  V_C
\end{equation}
where $V_C \subset V$ is a set of constrained vertices and $h:M\longrightarrow \mathbb{R}$ is a known function. The cardinality of the set $V_C$ must be at least $1$ in order for system (\ref{pois}) to have a unique solution. With the appropriate choice of $h$ we can use the Poisson equation for our purpose. Indeed, if we choose the function $h$ as suggested by Dong \textit{et al} in ~\cite{dong2005harmonic} then the solution $f$ of the Poisson equation gives us a scalar field whose level sets follow one of the principal curvatures of the underlying manifold. Specifically, this can be done by solving:
\begin{equation}
\label{mysystem}
\Delta f (v)=\kappa(v) \textrm{ where }  f(v_{source})=c
\end{equation} 
where $\kappa(v)$ is the mean curvature at the vertex $v$:
\begin{equation}
\kappa(v_i)=\frac{1}{4A_{mixed}(v_i)}\sum_{j \in N(i)}(\cot \theta_{ij}+\cot \beta_{ij})||(v_i-v_j)||
\end{equation}
here the angles $\theta_{ij}$ and $\beta_{ij} $ are given in Figure \ref{ang},

and $A_{mixed}(v_i)$ is the surface mixed area around the vertex $v_i$   ~\cite{meyer2003discrete}. Examples of Poisson fields on triangulated meshes are shown in \ref{poisson123}.

\section{Conclusion and Future Work}

The parallelization of topological data analysis algorithms is still in its infancy. There are plenty of existing topological machineries, such as Morse theory, that offer a plethora of tools that can be utilized to obtain robust and efficient parallel algorithms. In this paper, we presented a work that utilizes Morse theory to obtain a parallel algorithm for augmented Reeb graphs.

The parallel algorithm that we present here has elements that makes it generalizable to a Reeb graph algorithm on a general simplicial complex. However, we thought that this would make the discussion more complicated in many parts of the algorithm. We plan to pursue this direction in future work.

\section*{Acknowledgements}
This work was supported in part by the National Science Foundation (IIS-1513616 and IIS-1845204).

\bibliographystyle{abbrv}

\bibliography{refs}

\end{document}